\documentclass[fleqn,usenatbib]{mnras}

\usepackage[T1]{fontenc}
\usepackage{ae,aecompl}

\DeclareRobustCommand{\VAN}[3]{#2}
\let\VANthebibliography\thebibliography
\def\thebibliography{\DeclareRobustCommand{\VAN}[3]{##3}\VANthebibliography}

\usepackage{graphicx}	
\usepackage{amsmath}	
\usepackage{float}

\usepackage{newtxtext,newtxmath}

\title[GWsim]{GWsim: a code to simulate gravitational waves propagating in a potential well}

\author[Jian-hua He]{
Jian-hua He,$^{1,2}$\thanks{E-mail: hejianhua@nju.edu.cn}
\\
$^{1}$School of Astronomy and Space Science, Nanjing University, Nanjing 210023, P. R. China\\
$^{2}$Key Laboratory of Modern Astronomy and Astrophysics (Nanjing University), Ministry of
Education, Nanjing 210023, China\\
\\
}

\date{Accepted XXX. Received YYY; in original form ZZZ}

\pubyear{2021}

\begin{document}
\label{firstpage}
\pagerange{\pageref{firstpage}--\pageref{lastpage}}
\maketitle

\begin{abstract}
We present a code to simulate the propagation of GWs in a potential well in the time domain. Our code uses the finite element method (FEM) based on the publicly available code {\it deal.ii}. We test our code using a point source monochromatic spherical wave. We examine not only the waveform observed by a local observer but also the global energy conservation of the waves. We find that our numerical results agree with the analytical predictions very well. Based on our code, we study the propagation of the leading wavefront of GWs in a potential well. We find that our numerical results agree with the results obtained from tracing null geodesics very well. Based on our simulations, we also test the accuracy of the thin-lens model in predicting the positions of the wavefront. We find that the analytical formula of the Shapiro-time delay is only accurate in regimes that are far away from the center of the potential well. However, near the optic axis, the analytical formula shows significant differences from the simulated ones. Besides these results, we find that unlike the conventional images in geometric optics, GWs can not be sheltered by the scatterer due to wave effects. The signals of GWs can circle around the scatterer and travel along the optic axis until arrive at a distant observer, which is an important observational consequence in such a system.
\end{abstract}

\begin{keywords}
gravitational waves -- wave effects -- wavefront effects
\end{keywords}



\section{Introduction}

The landmark discovery of gravitational waves~\cite{Abbott:2016blz}, ripples in the fabric of spacetime triggered by the merger of two black holes, confirms a major prediction of Einstein's general relativity. It marks the climax of century-long speculation and decades-long painstaking work for hunting such waves. The direct detection of gravitational waves ushered us into a new era of astronomy, allowing us to probe the Universe in an unprecedented way. 

In the coming decades, ground- and space-based GW experiments will continue to study GWs, such as the Einstein Telescope (ET)~\cite{Einstein_tele}, 40-km LIGO~\cite{LIGO40}, eLISA~\cite{eLISA}, DECIGO~\cite{Sato_2009}, and Pulsar Timing Arrays (PTA)~\cite{2010CQGra..27h4013H}. The range of the frequency of GWs will be explored from ultra-low frequency regimes such as $10^{-9} {\rm Hz}-10^{-8} {\rm Hz}$ (PTA), $10^{-4}{\rm Hz}-10^{-1} {\rm Hz}$ (eLISA) ,and $10^{-2}{\rm Hz}-1 {\rm Hz}$ (DECIGO) to high frequency regimes such as $10{\rm Hz}-10^{4} {\rm Hz}$ (ET and 40-km LIGO). These experiments will usher us into an era of routine GW astronomy and enable us to study the GW phenomenon in unprecedented detail. The combination of these experiments promises to address a variety of outstanding problems in astrophysics, cosmology, and particle physics. 

The low-frequency GWs observed by space-borne GW interferometer may shed light on the population of supermassive BHs (SMBHs) with a mass $>10^6M_{\odot}$, which is believed to ubiquity reside in the centers of galaxies~\cite{Bellovary:2015ifg,Bartos:2016dgn,Stone:2016wzz}. SMBHs are major cosmic players and are important to galaxy formation~\cite{Haehnelt:2000sx}. The associated jets and accretion disks are the most energetic phenomena in the Universe, which often outshine the entire host galaxy. Therefore, their properties play an important role in the dynamics of stars and gas at galactic scales~\cite{Ferrarese:2000se}. However, the details of the birth and growth of SMBHs are not well understood yet. The observation of low-frequency GWs, such as in the millihertz band, can probe the spacetime geometry of black hole binaries and their surroundings, which opens a window that can inform us of the properties of these BHs throughout cosmic history. GW observations, thus, can help us to answer many fundamental questions, as to, where and when do the first massive black holes form, how do they grow and assemble, and how their properties are connected to the properties of their host galaxies?

However, in order to achieve these scientific goals, it is crucial to accurately infer the physical properties of the distant source from the observed GW signals. This process, however, is likely beset by the wave nature of GWs. Unlike the visible light, the wavelength of the low-frequency GWs is much larger. In the millihertz band, for example, the wavelength can be as large as several times the astronomical unit (${\rm AU}$). When the wavelength of the GWs is comparable to the Schwarzschild radius of an object, its propagation no longer obeys the geometrical optics. Further, if the gravitational waves come from the same compact binaries, they should be coherent and interference. In all these cases, the wave effects of GWs are significant and have to be taken into account~\cite{Bontz,Ohanian:1974ys}. 

The wave effects in a lensing system have already been studied in the previous work~\cite{PhysRevD.34.1708,Deguchi,Schneider,Ruffa_1999,DePaolis:2002tw,Takahashi:2003ix,1999PThPS.133..137N,Suyama:2005mx,Christian:2018vsi,Zakharov_2002,Liao:2019aqq,Macquart:2004sh,PhysRevLett.80.1138,Dai:2018enj,PhysRevD.90.062003,Yoo:2013cia,Nambu:2019sqn}. In these pioneer work (e.g.~\cite{Schneider}), the lensing system is usually considered as a thin lens model: the gravitational field of the compact lens is weak, and the deflection angle due to the lens masses is small. The impact parameter of the incident rays is thought to be much larger than the Schwarzschild radius of the lens mass.  As such, by construction the thin-lens model does not work for light rays that are close to the optic axis. Moreover, the changes in the amplitude of the incident waves are negligible at the location of the lens mass. The changes in the amplitude, however, are mainly due to the deflection near the lens, which is small at the beginning but becomes significant after the long journey of the light rays from the lens to the observer (see discussions in (4.86) in ~\cite{Schneider}). 

In addition to the thin-lens model, the previous work also applied {\it Kirchhoff's} diffraction theory to the gravitational lensing system~\cite{Schneider}. The lens is considered as an optic aperture. The lensed signal is then thought to be the diffraction of the waves from the aperture. An important assumption in {\it Kirchhoff's} theory is that boundaries outside the aperture are set to be zero. {\it Kirchhoff's} theory therefore assumes that only waves re-radiated from the aperture can be observed by the observer. The diffraction integral over the aperture, indeed, only considers the scattered waves, which neglects the contribution of the original incident waves (see Eq.(5) in ~\cite{Takahashi:2005sxa}). This is valid in the system of wave optics, as in this case the original incident waves are blocked by obstacles outside the aperture. However, if the amplitude of the scattered wave is comparable to that of the incident wave, the original incident waves have to be considered as well (This similar to the Babinet's principle in wave optics, in which the aperture is replaced by a screen. In this case, the incident and scattered waves have to be considered simultaneously.) 

If we consider the incident waves, the lensing system can be treated as a scattering  system~\cite{GR_steven_weinberg,Peters,Takahashi:2005sxa,Sorge:2015yoa}. Indeed, the scattering of waves by potential wells has been well studied in Quantum Mechanics. As shown in ~\cite{Peters}, the wave equation of GWs propagating in a non-uniform background spacetime is similar to that of the well-known Schr{\"o}dinger's equation for the scattering of two charged particles. Moreover, ~\cite{Peters,Takahashi:2005sxa} also generalized the scattering model that uses a plane wave as the incident wave to a spherical one. 

Nevertheless, despite these progresses, there are still some concerns. The previous analyses mainly focus on the frequency-domain with implicit assumptions, such as that GWs are in the steady-state (e.g. \cite{Peters}) or stationary state (e.g.~\cite{Suyama:2005mx}). These analyses also assume that the wave zone of GWs is infinite, which neglects the causality and wavefront effects. However, the realistic GW signals are time-finite and, indeed, neither in a steady-state nor stationary state. Moreover, unlike in the 2D case (or even dimensions), a prominent feature of 3D (or odd dimensions) waves is that they follow the strong Huygens' principle. In free space, a perturbation in 3D space at a point $\vec{x}$ is visible at another point $\vec{x}'$ exactly at the time $t = |\vec{x} - \vec{x}'|/c$ but not later. As a result, a time-finite wave signal has a definite wave zone with clear leading and trailing wavefronts as boundaries. As the leading wavefront represents the transport of energy from one place to another, it has to follow the constraint of causality. The behavior of the leading wavefront, therefore, is different from the rest part of the waves. Since GWs have a finite wave zone, the wavefront effects should be taken into account. However, despite their importance, they have not yet been properly addressed in the literature. 

On the other hand, to effectively address these issues is, indeed, highly non-trivial. First, the leading wavefront is not necessarily to be continuous but can be a step-like function with a sharp edge (e.g. square waves). In this case, the sums of the Fourier series do not converge near the point of discontinuity of a piecewise smooth function. Even with a large number of the Fourier series, the sums oscillate at the discontinuity point, which is approximately $9\%$ of the jump of the discontinuity\footnote{In mathematics, the Fourier series converges only in the sense of $L_2$ norm, which means that they converge up to functions with differences on a set of Lebesgue measure zero. However, they do not necessarily converge in the pointwise sense, unless the function is continuously differentiable.}. This phenomenon is known as the Gibbs phenomenon. Second, the wavefront may have a complex shape in a 3D potential well, which can lead to complex boundaries of the wave zones. The domain of the wave zone might even not be convex\footnote{The solutions of partial differential equations (PDEs) are strongly dependent on the shape of the boundaries of the domain. It is offten assumed that the shape of boundaries is regular or of Lipschitz continuity. As such, the domains of PDEs are usually assumed to be bounded convex domains.}. However, even for the simplest Laplacian operator, if the dimension of space is greater than 1, the eigenvalues and eigenfunctions are known only for simple geometries (see detailed discussions in chapter 1 of \cite{grossmann2007numerical}). As such, in general cases, it is difficult to analyze the behaviors of the wavefront in a potential well using the spectrum method based on eigenfunctions (e.g. Fourier analysis). Therefore, numerical techniques are called for in such situations.

This work aims to develop a code that can efficiently simulate the propagation of GWs in a potential well. The main technique used in this work is called the finite element method (FEM). The FEM is an effective and well-developed method for numerically solving partial differential equations (PDEs) (see e.g. text book~\cite{FEMbook} for details). Unlike most of the conventional methods, such as the finite difference method (FDM), the FEM is based on the weak formulation of PDEs. The basic idea of the FEM is to present the domain of interest as an assembly of finite elements. On each finite element, the solution of PDEs is approximated by local shape functions. A continuous PDE problem can then be transformed into a discretized finite element with unknown nodal values. These unknowns form a system of linear algebraic equations, which can be solved numerically. Compared to the conventional numerical methods, such as the FDM, which requires regular meshes, the advantage of the FEM is that it can work for complex boundaries so that it can significantly minimize the effects of boundaries. Moreover, it can also provide very good precision even with simple approximation shape functions. Due to the locality of the approximation shape functions, the resulting linear algebraic equations are sparse, which can be solved efficiently in the numerical process.

Throughout this paper, we adopt the geometric unit $c=G=1$, in which $1\,{\rm Mpc}=1.02938\times 10^{14} {\rm Sec}$ and $1 M_{\odot}=4.92535\times 10^{-6} {\rm Sec}\,.$ The advantage of this unit system is that time and space have the same unit. 
 
This paper is organized as follows: In Section~\ref{sec:WF}, we introduce the Strong and Weak formulation of the wave equation. In Section~\ref{sec:FEM}, we describe the FEM for wave equations. In Section~\ref{sec:CT}, we present several tests of our code using spherical waves. In Section~\ref{sec:GWP}, we simulate GWs in a potential well. In Section~\ref{sec:cl}, we summarize and conclude this work.
\section{Strong and Weak formulation of the wave equation} \label{sec:WF}
In this section, we present an overview of the mathematical basis of the wave equation. To begin with, we discuss the {\it strong formulation} of the wave equation and then we will introduce the {\it weak formulation}.
\subsection{Strong formulation}
Let $\Omega \subset \mathbb{R}^3$ be a bounded domain with boundary $\partial \Omega$. The boundary value problem (BVP) we aim to study in this work can be presented as
\begin{equation}
c^2\nabla^2 u -\frac{\partial^2}{\partial t^2}u =c^2f\quad {\rm in} \quad \Omega\times (0,T] \label{strong}\,,
\end{equation}
which is a second-order linear hyperbolic partial differential equation (PDE) 
with initial conditions
\begin{align}
u(x,0) = u_0(x) \quad {\rm in} \quad \Omega  \,. \label{S1}
\end{align}
Let $\partial \Omega_1$ and $\partial \Omega_2$ are subsets of the boundary $\partial \Omega$ with $\partial \Omega_1\cup \partial \Omega_2=\partial \Omega$ and $\partial\Omega_1\cap\partial\Omega_2=\emptyset$. 
We assume that the boundary conditions on $\partial \Omega_1$ are Dirichlet and on $\partial \Omega_2$ is Neumann
\begin{align}
u(x,t) &= u_1(x,t) &{\rm on}& \quad \partial\Omega_1\times (0,T]  \label{S2}\\ 
\hat{n}\cdot\nabla u & = q(x,t) &{\rm on}& \quad \partial\Omega_2\times (0,T] \label{S3} \,, 
\end{align}
where $\hat{n}$ is the outward pointing unit vector normal to $\partial \Omega_2$. $q(x,t)$ is a given function on the boundary $\partial\Omega_2$.

In mathematics, Eq.~(\ref{strong}), together with the initial and boundary conditions Eqs.~(\ref{S1},\ref{S2},\ref{S3}), are called the {\it strong formulation}. The {\it strong formulation} is the most commonly used format of the wave equation. It can be numerically solved using some straightforward methods such as the finite difference method (FDM). However, the {\it strong formulation} is not the only format of the wave equation. In what follows, we introduce an alternative format, namely, the {\it weak formulation} of the above equations. 
\subsection{Weak formulation}
Multiply Eq.~(\ref{strong}) by {\it a test function} $\phi$ and then integrate over $\Omega$, we obtain a variational equation
\begin{align}
\begin{split}
&-\int_{\Omega}\nabla u \cdot \nabla (c^2\phi)\,\mathrm{d}x + \int_{\partial \Omega}(\hat{n}\cdot\nabla u) (c^2\phi)\,\mathrm{d}x -\frac{\partial^2}{\partial t^2}\int_{\Omega} u \phi \, \mathrm{d}x\\
=&\int_{\Omega} c^2f\phi \,\mathrm{d}x  \quad \forall \phi \in V 
\end{split}\,,\label{weak}
\end{align} 

where we have used the Green's formula
\begin{equation}
\int_{\Omega} c^2\phi \nabla^2 u \,\mathrm{d}x= -\int_{\Omega}\nabla(c^2\phi)\cdot\nabla u \,\mathrm{d}x + \int_{\partial \Omega}c^2\phi \hat{n}\cdot\nabla u \,\mathrm{d}x \,.\nonumber
\end{equation}
The second term on the right hand represents the boundary term, which can be further split into terms according to different boundary conditions
\begin{align}
	\begin{split}
\int_{\partial \Omega}(\hat{n}\cdot\nabla u) (c^2\phi)\,\mathrm{d}x &=\int_{\partial \Omega_1}(\hat{n}\cdot\nabla u) (c^2\phi)\,\mathrm{d}x \\
& + \int_{\partial \Omega_2}(\hat{n}\cdot\nabla u) (c^2\phi)\,\mathrm{d}x 
    \end{split}\,.
\end{align}

The test function $\phi$ is chosen in such a way that it vanishes on the subset of boundaries with Dirichlet boundary conditions ${\partial \Omega_1}$, which is equivalent to say that we restrict $\phi$ to lie in 
\begin{equation}
V:=\{\phi:\phi\in H^{1}(\Omega),\phi|_{\partial \Omega_1}=0\}\,,\label{Vspace}
\end{equation}
where $H^{1}(\Omega)=W^{1,2}(\Omega)$ is called the first order {\it Sobolev space}, which is also a Hilbert space, meaning that $\phi$ and its first order weak derivatives $\partial_x \phi$ are square integrable. 
\begin{equation}
\left\|\phi\right\|_{H^1(\Omega)}=\left[\int_{\Omega}\sum_{|\alpha|\leq 1} |\partial^{\alpha}_x\phi(x)|^2dx\right]^{\frac{1}{2}}<\infty\,.\nonumber
\end{equation}

Given the restrictions of the test function $\phi$, the boundary term in Eq.~(\ref{weak}) can be reduced to
\begin{align}
	\begin{split}
\int_{\partial \Omega}(\hat{n}\cdot\nabla u) (c^2\phi)\,\mathrm{d}x &=\int_{\partial \Omega_2}(\hat{n}\cdot\nabla u) (c^2\phi)  \\
&=-\int_{\partial \Omega_2}\frac{\partial{u}}{\partial{t}} (c\phi) \,\mathrm{d}x 
    \end{split} \,.
\end{align}
Here, in the second equality, we have imposed an absorbing boundary condition
\begin{equation}
\hat{n}\cdot\nabla u=-\frac{1}{c}\frac{\partial{u}}{\partial{t}}\quad {\rm on} \quad \partial\Omega_2\times (0,T] \,.\label{boundary2}
\end{equation}
The physical meaning of the absorbing boundary condition will be explained later on.

The above we have discussed the boundary condition on $\partial \Omega_2$. On the other hand, the boundary condition on $\partial \Omega_1$,
\begin{equation}
\left \{
\begin{aligned}
u(x,t) &= u_0(x,t)   \\
\frac{\partial u(x,t)}{\partial t} &= \frac{\partial u_0(x,t)}{\partial t}  
\end{aligned}
\right. \quad {\rm on} \quad \partial\Omega_1\times (0,T] \,, \label{boundary1}
\end{equation}
is called the {\it essential boundary condition}, which does not affect the variational
equation Eq.~(\ref{weak}) but this condition must be imposed directly on the solution itself.

Equipped with the boundary conditions Eqs.~(\ref{boundary2},\ref{boundary1}), Eq.~(\ref{weak}) is called the {\it weak formulation} of Eq.~(\ref{strong}). It can be shown that if Eq.~(\ref{weak}) holds for all $\phi\in V$, it has a unique solution $u\in V$~\cite{LaxMilg}. This solution is called the weak solution. It is obvious that the classical solutions of Eq.~(\ref{strong}) are also weak solutions. However, conversely, weak solutions must have sufficient smoothness to be the classical solutions, which dependents both on the shape of the boundary $\partial \Omega$ and the regularity of $c^2f$. For example, it can be shown that if $\Omega \subset \mathbb{R}^n$ has $C^k$ boundary (smooth up to $k$-th derivative) and $c^2f\in H^{k}(\Omega)$ with $k>\frac{n}{2}$, then $u\in C^2(\bar{\Omega})$, namely $u$ is a solution in the classical sense~\cite{nla:cat-vn1414651}. 

\subsection{Energy Conservation of the wave equation}
In addition to the {\it weak formulation} of the wave equation, another important property of the wave function Eq.~(\ref{strong}) is that it obeys the law of energy conservation. To see this point, we multiply Eq.~(\ref{strong}) by $\frac{\partial u}{\partial t}$ and integrate over the domain $\Omega$
\begin{align}
\begin{split}
&\frac{1}{2}\frac{\partial}{\partial t}\int_{\Omega}\left[\left(\frac{\partial{u}}{\partial t}\right)^2+c^2\left(\nabla u\right)^2\right] \,\mathrm{d}x \\
=& \int_{\partial\Omega}c^2\left(\frac{\partial u}{\partial t}\right)(\hat{n}\cdot\nabla u)\,\mathrm{d}x-\int_{\Omega}\left(\frac{\partial u}{\partial t}\right)\left[\left(\nabla c^2\right)\cdot\nabla u+c^2f\right]\,\mathrm{d}x 
\end{split}\,. \label{Energy_con}
\end{align}

Note that the total energy carried by a wave in the domain $\Omega$ is defined by
\begin{equation}
E(t)=\frac{1}{2}\int_{\Omega}\left[\left(\frac{\partial{u}}{\partial t}\right)^2+c^2\left(\nabla u\right)^2\right] \,\mathrm{d}x \,. \label{defenergy}
\end{equation}
The left hand of the equality of Eq.~(\ref{Energy_con}) then represents the change of the total energy of wave with time.
From Eq.~(\ref{Energy_con}), if the boundary condition on the right hand side of the equality is a constant ($\frac{\partial{u}}{\partial t}|_{\partial \Omega}=0$) and there is no force source $f=0$ in the second term, for a constant speed of wave $\nabla c^2=0$, the total energy is conserved $\frac{dE(t)}{dt}=0$.

Equation~(\ref{Energy_con}) is an important feature of the wave function. We shall discuss this point in our following numerical analyses.

\section{The Finite Element Method } \label{sec:FEM}
Before presenting our numerical analyses, we introduce an auxiliary function $v:=\frac{\partial u}{\partial t}$ and adopt the common notation 
\begin{equation}
\left(f,g\right)_{\Omega}=\int_{\Omega}f(x)g(x)\,\mathrm{d}x\quad,
\end{equation}
for convenience.
Equation~(\ref{weak}) in this case can be re-written as a set of two equations
\begin{equation}
 \left.\begin{aligned}
        \left(\frac{\partial u}{\partial t},\phi\right)_{\Omega}-\left(v,\phi\right)_{\Omega}&=0  \\
\left(\nabla u,\nabla(c^2\phi)\right)_{\Omega}+\left(\frac{\partial u}{\partial t},c\phi\right)_{\partial \Omega}+\left(\frac{\partial v}{\partial t},\phi \right)_{\Omega}+\left(c^2f,\phi\right)_{\Omega}&=0
       \end{aligned}
 \right \} \quad \forall \phi \in V \label{evtwo} \,.
\end{equation}
To solve the above equations numerically, we need to discretize these equations first. In general for a time-dependent problem, the discretization can be done either for the spatial variables first or the temporal variables first. The method to discretize the spatial variables first is called the (vertical) method of lines (MOL). Once the spatial variables are discretized, each variable needs to be solved evolving with time. This constitutes a large system of ordinary differential equations (ODEs). The advantage of this method is that ODEs can be numerically solved by well-developed ODE solvers that are efficient and of high-order accuracy(see discussions in Chapter 5 of Ref.~\cite{grossmann2007numerical}). 

The alternative approach is called the {\it Rothe's method}, in which the temporal variable is discretized first. The advantage of this method is that the spatial resolution can be changed at each time step, which allows for the usage of the technique of {\it adaptive mesh refinement} (AMR). However, when trying to apply high-order temporal discretization schemes, {\it Rothe's method} is rather awkward.

Whether the time variable or the spatial variable should be discretized first has long been debated in the numerical analysis community. In this work, however, we follow the {\it Rothe's method} as we shall show that the spatial resolution is, indeed, the major limiting factor for the problem that we are interested in. The {\it Rothe's method} allows for great freedom to deal with the spatial resolution.

\subsection{discretization in time}
We use superscript $n$ to indicate the number of a time step and $k=t_{n}-t_{n-1}$ is the length of the present time step. We discretize the time derivative as 
\begin{align}
	\left \{ \begin{aligned}
\frac{\partial v}{\partial t} &\approx \frac{v^{n}-v^{n-1}}{k}\quad \\
\frac{\partial u}{\partial t} &\approx \frac{u^{n}-u^{n-1}}{k}
\end{aligned}
 \right. \,.
\end{align}

The discretization scheme we adopted here also involves the spatial quantities at two different time steps, which is known as the $\theta$-scheme. Equations~(\ref{evtwo}) then can be represented as 

\begin{equation}
\left. \begin{aligned}
&\left(\frac{u^n-u^{n-1}}{k},\phi\right)_{\Omega}-\left(\theta v^n+(1-\theta)v^{n-1},\phi\right)_{\Omega} =0\,; \\
&-\left(\nabla [\theta u^n+(1-\theta)u^{n-1}],\nabla(c^2\phi)\right)_{\Omega}
-\left(\frac{u^n-u^{n-1}}{k},c\phi\right)_{\partial \Omega} 
\\
=&\left(\frac{v^n-v^{n-1}}{k},\phi \right)_{\Omega}+\left(c^2
[\theta f^n+(1-\theta)f^{n-1}],\phi\right)_{\Omega} 
\end{aligned} \right\}  \quad \forall \phi \in V \,. \label{Time_discretization}
\end{equation}
When $\theta = 0$, the scheme is called the forward or explicit Euler method. If $\theta = 1$, it reduces to the backward or implicit Euler method. The Euler method, no matter implicit or explicit, is only of first-order accurate. Therefore we do not adopt here.

The scheme we use here is called the {\it Crank-Nicolson Scheme}, namely $\theta = 1/2$, which uses the midpoint between two different time steps. This scheme is of second-order accuracy. As the midpoint rule is symplectic, the most important feature of this scheme is that it is energy-preserving. We shall discuss this point in our following numerical analyses.

\subsection{discretization in space}
We discretize the domain $\Omega \subset \mathbb{R}^3$ using the Finite Elements Methods (FEM). The FEM is an effective method for numerically solving partial differential equations. There are many excellent textbooks on this topic (Readers are referred to e.g. Ref.~\cite{FEMbook} for more details). Here, we will not go into too many mathematical details. Instead, we only summarise the main steps here.

First, the domain $\Omega$ is decomposed into subdomains $\Omega_i$, which consist of rectangles and triangles. This is called {\it decomposition} or {\it triangulation}. The vertices of rectangles and triangles in the domain $\Omega$ are called mesh points or nodes. Let $\Omega_h$ denote the set of all nodes of the decomposition. On each node, we construct a test function $\phi_i\in V\,,i=1,..,N$, where $V$ is the space defined in Eq.~(\ref{Vspace}) and $N$ is the total number of nodes in the domain. The test function $\phi_i$ is required to have the property
\begin{equation}
\phi_i(p^k)=\delta_{ik}, \quad i,k=1,..,N, \quad p^k\in \Omega_h \,, \nonumber
\end{equation}
where 
\begin{equation}
\delta_{ik} = \left \{
\begin{aligned}
1 \quad &{\rm for}& \quad i=k \\
0 \quad &{\rm for}& \quad i\ne k \nonumber
\end{aligned}
\right. \quad.
\end{equation}

Thus, $\phi_i$ has non-zero values only on the node with $k=i$ and its adjacent subdomains. It vanishes on other parts of the domain $\Omega$. The test function $\phi_i$ constructed this way is called the {\it shape function}. Clearly, $\phi_i\in V$ on different nodes are linearly independent. We denote the space spanned by $\phi_i$ as
$V_h:=\mathrm{span}\{\phi_i\}_{i=1}^N$, which is a subspace of $V$. 

Next, we expand the scalar fields $u,v,c^2f$ in terms of $\phi_i$ 
\begin{equation}
u_h = \sum_{i=1}^N u_i \phi_i \,,
v_h = \sum_{i=1}^N v_i \phi_i \,,
c^2f_h = \sum_{i=1}^N c^2f_i \phi_i \,,\label{expansion}
\end{equation}
where $u_h\,,v_h\,,c^2f_h\in V_h$. These functions are approximations of the original scalar fields.
The advantage of this expansion is that the coefficients are precisely the values of these functions on the node points (e.g. $u_i=u_h(p_i)$).

Inserting the above expansions into Eqs.~(\ref{Time_discretization}) and noting that Eqs.~(\ref{Time_discretization}) holds for any $\phi\in V$, thus, we can choose $\phi$ as $\phi=\phi_j$, where $j=1,..,N$. Then we obtain $N$ different equations, which form a linear system
\begin{align} 
[k\theta(A+D)+B]U^n&=-MV^n+G_2 \nonumber\\
&-k[\theta F^n+(1-\theta)F^{n-1}] \,,\label{U1}\\
MU^n&=k\theta MV^n+G_1 \label{V1}\,,
\end{align}
where $G_1$ and $G_2$ are defined by
\begin{equation}
\left \{
\begin{aligned} 
G_1&=MU^{n-1}+k(1-\theta)MV^{n-1}\\
G_2&=[-k(1-\theta)(A+D)+B]U^{n-1}+MV^{n-1}
\end{aligned}
\right.\,, \nonumber
\end{equation}
and the elements of the matrixes are defined by
\begin{equation}
\left \{
\begin{aligned} 
D_{ij}&=\left(\nabla (c^2) \phi_i,\nabla \phi_j\right)_{\Omega} \\
A_{ij}&=\left(c^2 \nabla \phi_i,\nabla \phi_j\right)_{\Omega}\\
M_{ij}&=\left(\phi_i,\phi_j\right)_{\Omega}\\
B_{ij}&=\left(c \phi_i, \phi_j\right)_{\partial \Omega}\\
F^n_{i}&=\left(c^2f^n, \phi_i\right)_{\Omega}\\
U^n_{i}&=\left(u^n, \phi_i\right)_{\Omega}\\
V^n_{i}&=\left(v^n, \phi_i\right)_{\Omega}
\end{aligned}
\right.\,. \nonumber
\end{equation}
It is worth noting that the wave speed $c$ is not a constant so that the matrix $D$ is not symmetric $D_{ij}\neq D_{ji}$, which needs to be kept consistent with Eqs.~(\ref{expansion}).
Equations~(\ref{U1},\ref{V1}) are called the {\it Galerkin equations}, from which the unknown coefficients $u_i$
and $v_i$ can be solved. Note that Eq.~(\ref{U1}) explicitly contains $V^n$, which is unknown at the time step $n$. However, we can multiply Eq.~(\ref{U1}) by $k\theta$ and then use Eq.~(\ref{V1}) to eliminate $V^n$. Then we obtain 
\begin{align}
[M+k^2\theta^2(A+D)+k\theta B]U^n &  =  G_1+k\theta G_2  \nonumber\\
 &-  k^2\theta[\theta F^n+(1-\theta)F^{n-1}] \,,\label{U2}
\end{align}
for $U^n$ and 
\begin{align}
MV^n&=-[k \theta (A+D)+B]U^n+G_2 \nonumber\\
 &-  k[\theta F^n+(1-\theta)F^{n-1}] \,,\label{V2}
\end{align}
for $V^n$.

Now, $U^n$ can be solved using the information on the time step $n-1$. Then $V^n$ can be solved with the knowledge of $U^n$. In practice, Eqs.~(\ref{U2},\ref{V2}) usually contain a very large number of equations. The {\it rank} of Eqs.~(\ref{U2},\ref{V2}) (the number of linear independent equations) is called the degrees-of-freedom (DOF) of the system, which depends both on the number of node points and the freedoms in the {\it shape functions}.  

\subsection{linear solvers}
As the systems of linear Eqs.~(\ref{U2},\ref{V2}) may contain a large number of unknowns(e.g. which can be easily up to a few $10^{7}$), direct methods such as the LU decomposition is inefficient in this case. One has to use iterative methods such as the Conjugate Gradient (CG) method, which is efficient for large but sparse matrices. However, our cases are complicated due to the varying wave speed $c$. The matrix $D$ is no longer symmetric. As the CG method can be only applied to symmetric, positive-definite matrices, we can not use the CG method in this work. Instead, we adopt the GMRES (a generalized minimal residual algorithm for solving non-symmetric linear systems) method~\cite{osti_409863}, which does not require any specific properties of the matrices.  

\subsection{numerical implementation}
In this work, we use the public available code {\bf deal.ii}~\cite{dealII91,BangerthHartmannKanschat2007,dealII90} to numerically solve Eqs.~(\ref{U2},\ref{V2}). {\bf deal.ii} is a C++ program library aimed at numerically solving partial differential equations using modern finite element method. Coupled to stand-alone linear algebra libraries, such as PETSc~\cite{abhyankar2018petsc,petsc-web-page,petsc-user-ref,petsc-efficient},  {\bf deal.ii} supports massively parallel computing of vary large linear systems of equations. {\bf deal.ii} also provides convenient tools for {\it triangulation} of various geometries of the simulation domain. Therefore, it is straightforward to implement Eqs.~(\ref{U2},\ref{V2}) in the {\bf deal.ii} code. The detailed numerical analyses are presented in the next section.
 
\section{Code Tests} \label{sec:CT}
In this section, we present several tests of our code using an important physical model that describes waves emitting from a point source. Moreover, we shall discuss the {\it Huygens-Fresnel principle} for wave functions. We use the {\it Huygens-Fresnel principle} to set boundary conditions so that we can avoid mathematical singularities of spherical waves at the origin.
\subsection{Spherical Waves from a Point Source and the Huygens-Fresnel principle}
The boundary-initial value problem for the propagation of waves from a point source can be presented as
\begin{equation}
\nabla^2 u -\frac{1}{c^2}\frac{\partial^2}{\partial t^2}u =-Q(t)\delta(x) \quad {\rm in} \quad \Omega\times (0,T] \,, \label{pointsource}
\end{equation}
where $Q(t)$ is the time-dependent waveform of the point source and $\delta$ is the Dirac delta function. The wave equation has a fundamental solution, which is the distributional solution of
\begin{equation}
\nabla^2 G -\frac{1}{c^2}\frac{\partial^2}{\partial t^2}G =-\delta(t)\delta(x) \,.
\end{equation}
The distribution $G$ is also called the Green's function. In free-space, the explicit form of $G$ for an outgoing wave is given by  
\begin{equation}
	G^{+}(x,t;x',t')=\frac{\delta(t'-[t-\frac{\left\|x-x'\right\|}{c}])}{4\pi\left\|x-x'\right\|}\,.
\end{equation}
The above equation is known as the {\it retarded Green's function} in free-space. The general solution of Eq.~(\ref{pointsource}) then has the form
\begin{align}
	\begin{split}
u(x,t) &= \int\int G^{+}(x,t;x',t')Q(t')\delta(x')dx'dt' \\
&= \frac{Q(t-\frac{r}{c})}{4\pi r} \,, \label{retarded_potential}
    \end{split}
\end{align}
where $r$ is the norm of $x$, namely, $r=\left\|x\right\|$.

Although the wave function from a point source has a very simple analytic expression Eq.~(\ref{retarded_potential}), numerically solving Eq.~(\ref{pointsource}) is, indeed, non-trivial.
A challenge is that the point source on the right-hand side of Eq.~(\ref{pointsource}) is singular, which may cause numerical problems. In practice, in order to avoid this problem, we can make use of the Huygens-Fresnel principle to numerically solve Eq.~(\ref{pointsource}) on a slightly different domain without the origin but with new boundaries
\begin{align}
\nabla^2 u -\frac{1}{c^2}\frac{\partial^2}{\partial t^2}u &=0 \quad {\rm in} \quad \Omega/\{0\}\times (0,T]\,,
\end{align}
The Huygens-Fresnel principle states that the original waves from a source propagating to a distant observer can be considered as re-radiating from a wavefront of the original wave rather than directly from the source. As such, we can use the wavefront of the original wave at some radius as a new boundary for the wave equation. In this way, we can avoid the singularity in Eq.~(\ref{pointsource}) at the origin. A mathematical description of the Huygens-Fresnel principle for spherical waves is provided in Appendix~\ref{Appdex:sw}. 

\begin{figure}
{\includegraphics[width=\linewidth]{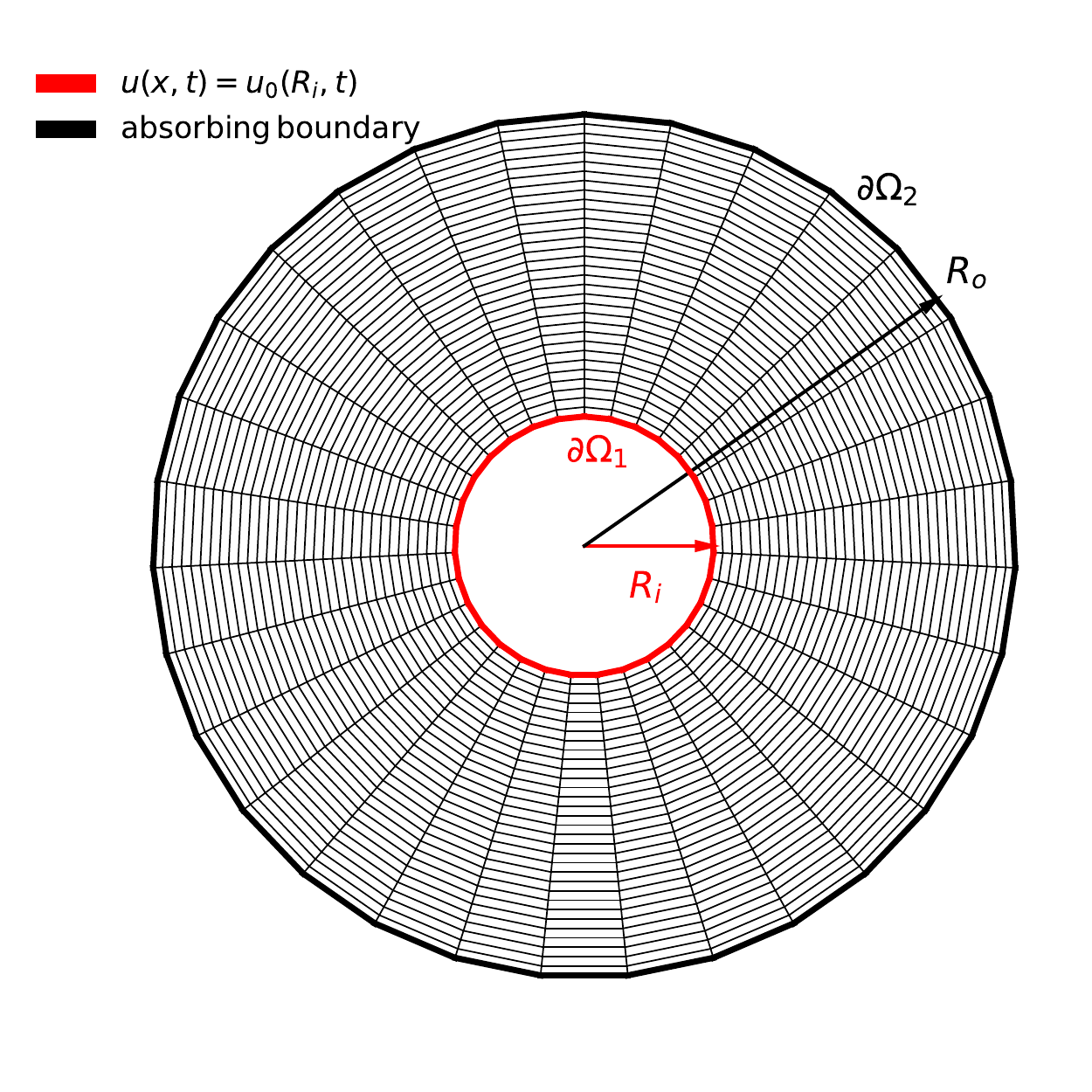}}
\caption{The boundaries and triangulation of our simulation domain. The domain consists of the bulk regions between the two spherical shells. The plot shows a slice taken through the center of the sphere. The meshes have a resolution of $2^5$. On $\partial \Omega_1$, we set the boundary conditions as the wavefront of the original wave from the point source propagating there. However, we set the absorbing boundary condition on $\partial \Omega_2$. \label{spherical_domain}}
\end{figure}

Figure~\ref{spherical_domain} shows the new boundaries and triangulation of our simulation domain. The domain consists of the bulk regions between the two spherical shells with an inner radius $R_i$ and an outer radius $R_o$, respectively. We set the boundary conditions at the inner shell $R_i$ as
\begin{equation}
u(R_i,t)=\left \{
	\begin{aligned} 
	&\frac{Q(t-\frac{R_i}{c})}{4\pi R_i} \quad &t\ge R_i/c\\
	&0\quad &t<R_i/c
	\end{aligned}
	\right.\quad {\rm on} \quad \partial \Omega_1\times (0,T]\,, \nonumber\\
\end{equation}
which is exactly the wavefront of the original point source propagating there. At the outer shell, we adopt the absorbing boundary condition
\begin{equation}
\frac{\partial u}{\partial r}(R_o,t) = - \frac{1}{c}\frac{\partial u}{\partial t} \quad {\rm on} \quad \partial \Omega_2\times (0,T]\,. \label{absorbing}
\end{equation}
The absorbing boundary condition is also called the non-reflecting boundary conditions or radiating boundary conditions. These boundary conditions can absorb and eliminate the reflections of waves on the boundaries. Therefore, with these boundary conditions, we can approximate the propagation of waves in free-space using a limited volume of the simulation domain. In addition to the boundary conditions, we set the initial conditions as
\begin{equation}
	u(x,0) = 0\quad {\rm in} \quad \Omega\,.
\end{equation}

After fixing the boundary and initial conditions, we perform several numerical tests. In these tests, we choose a concrete waveform $Q(t)=-A\sin(\omega t)$. Thus, the spherical wave has an analytical expression
\begin{equation}
u(x,t)=-\frac{A\sin[\omega (t-r/c)]}{4\pi r}\,, \label{waveanalytic}
\end{equation}
where $\omega = 2\pi f\,$ and $f$ is the frequency of waves. $A$ is the amplitude of the wave. In our test, we set $R_o=10^7M_{\odot}=49.2535{\rm Sec}\,,$$R_i=9\times10^6M_{\odot}=44.32815{\rm Sec}$ and $f=1$. Further, we set $A=4\pi R_i$, namely, the amplitude of the waves are normalized to unity at the inner boundary.

\subsection{The choice of time step}
To perform our simulations, we further need to set the time step. For a time-dependent wave equation, an {\it explicit} discretization method is only stable if the time step is small enough so that the wave can have enough time to travel through the space discretization (see. Chapter 2 in Ref.~\cite{grossmann2007numerical}). This condition is known as the Courant-Friedrichs-Levi condition

\begin{equation}
	k\le\frac{\sigma}{c}\,,
\end{equation}
where $\sigma$ is the size of the mesh and $k$ is the size of the time step as mentioned before. However, for the {\it implicit} methods such as the one used in this work, it can be stable for arbitrary step sizes if the scheme is {\it upwind} (see. Chapter 2 in Ref.~\cite{grossmann2007numerical}). However, a stable scheme is not sufficient to guarantee a correct solution to the wave equation. This is because of the oscillating features of waves. If the frequency of waves is too high, the wavelength will be too small and it is difficult to resolve the waveform within one period. The shape of the waves, in this case, is not smooth relative to the size of the meshes, which can lead to large errors in the numerical simulations. Therefore, in order to get a correct solution, the time step and the size of the meshes should be small enough so that the waveform within one period can be well resolved. As such, in our simulations, we set $k=\lambda/30$ and the size of the mesh $\sigma<\lambda/7$. We have tested that further reducing the size of the mesh $\sigma$ does not improve the results appreciably.

\begin{figure}
{\includegraphics[width=\linewidth]{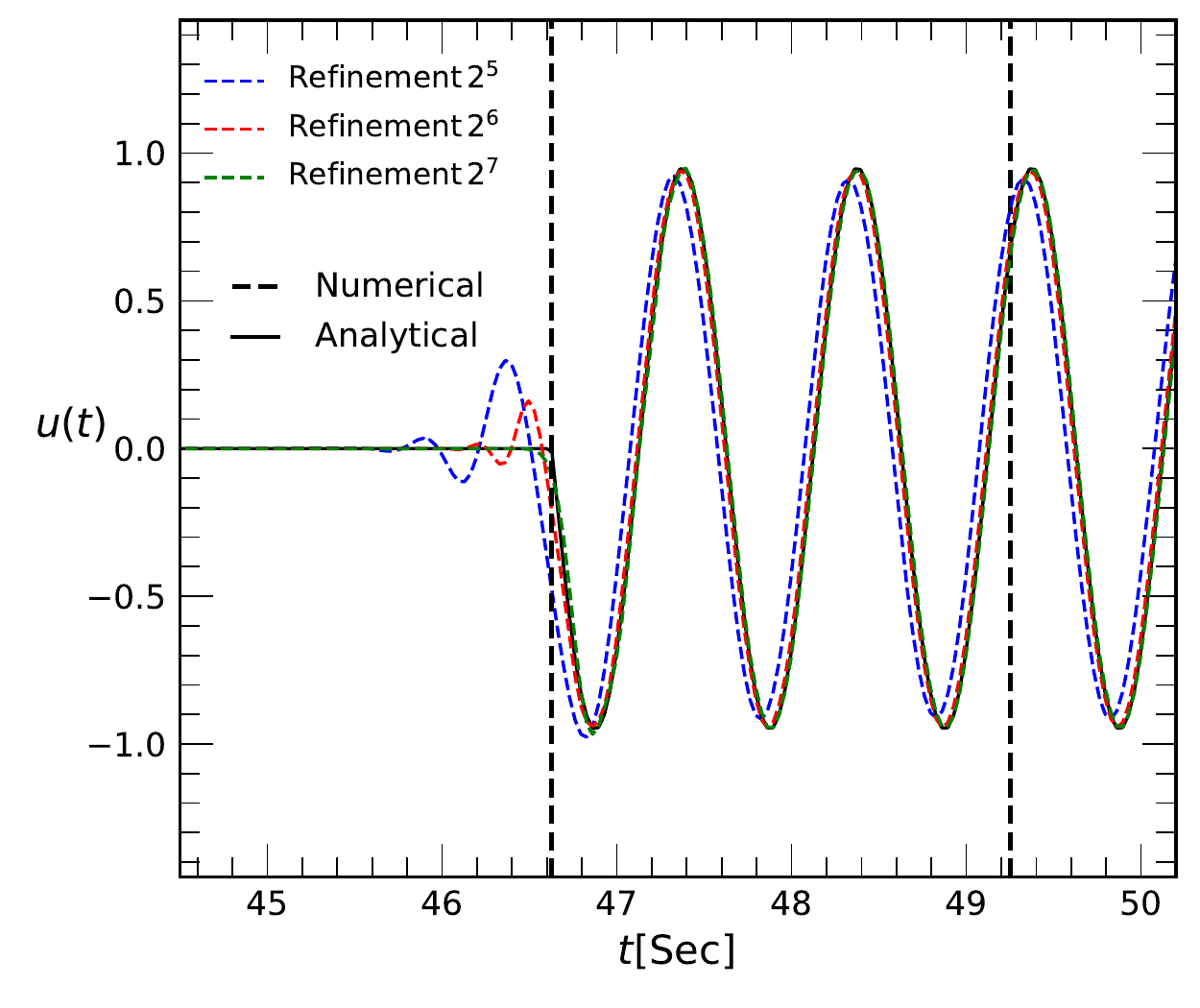}}
\caption{Numerical results (dashed lines) against analytical results (solid lines).
The waveform is observed by an observer located at $r=46.63{\rm Sec}$. The left vertical dashed line indicates the epoch that the wavefront just arrives at the observer. The right vertical dashed line indicates the epoch that the wavefront arrives at the outer boundary of our simulation domain. After this epoch, the wave starts getting out of the simulation domain. Dashed lines with different colors show the results obtained using different spatial resolutions.\label{SphericalWave}}
\end{figure}

\subsection{Numerical results}
We perform a suite of simulations with different spatial resolutions. In these simulations, we adopt the linear Lagrange test function, which depends only on the values of the finite elements on their vertices but does not involve any derivatives of the unknown fields. Therefore the degree of freedom (DOF) associated with each vertex is one. In this case, the total DOF is simply the total number of the expansion coefficients of the unknown field ($u_i$ or $v_i$). This number is also equal to the total number of linear algebraic equations in the system. Table~\ref{Table_Re} lists the level of refinement and the corresponding DOF.

\begin{table}
\centering
\caption{The degree of freedom of FEM for different resolutions \label{Table_Re}}
\begin{tabular}{c|c}
\hline 
Refinement & DOF \\ 
\hline 
$2^5$ & 202818 \\ 
\hline 
$2^6$ & 1597570 \\ 
\hline 
$2^7$ & 12681474 \\ 
\hline 
\end{tabular} 
\end{table}

Figure~\ref{SphericalWave} shows the waveform observed by an observer located at $r=46.63{\rm Sec}$. Dashed lines represent the numerical results and the solid lines are obtained from Eq.~(\ref{waveanalytic}).
The dashed vertical line indicates the epoch that the wavefront just arrives at the observer. Dashed lines with different colors show the results with different resolutions. With high spatial resolution $2^7$ (blue dashed line), the numerical results agree with the analytic one very well. This is expected as the accuracy of the FEM is strongly dependent on the resolution of elements.

After investigating the time-domain waveform for a particular local observer, we consider the integral property of the wave equation. Inserting Eq.~(\ref{waveanalytic}) into Eq.~(\ref{defenergy}), we obtain the total energy of waves in the domain
\begin{align}
\begin{split}
E(t)=&\pi R_i\left\{\omega R_i \sin[2\omega(t-R_i)]
-\cos[2\omega(t-R_i)]+1\right\}\\
+&2\pi\omega^2R_i^2(t-R_i), \quad t\in[R_i,R_o]\,. \label{totalenergy}
\end{split}
\end{align}
Then we measure the total energy in our simulation numerically. 

\begin{figure}
{\includegraphics[width=\linewidth]{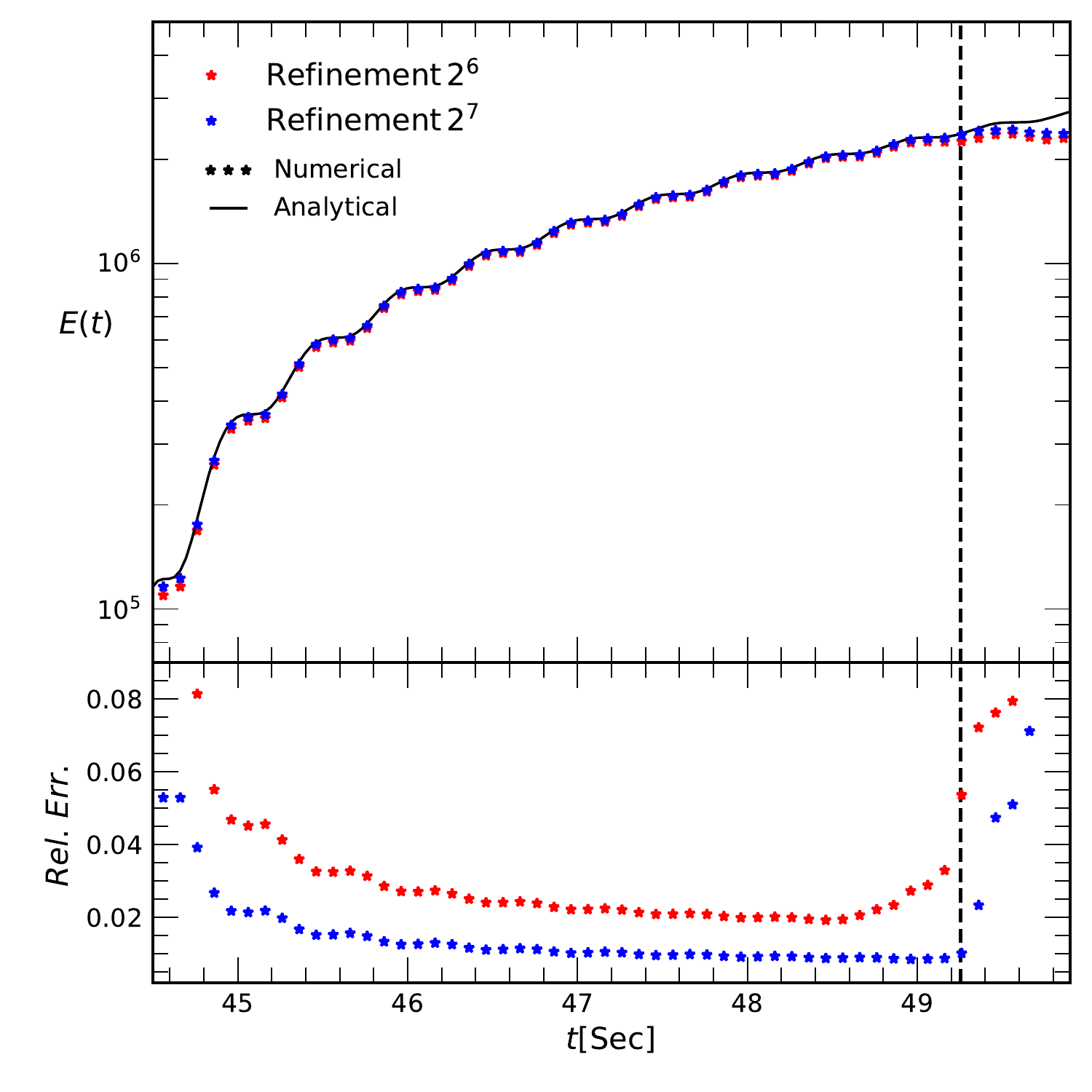}}
\caption{Upper panel: Comparisons of the total energy between numerical simulations (stars) and the exact analytic solution (solid line). Red and blue colors are for spatial resolution $2^6$ and $2^7$, respectively. The dashed vertical line indicates the epoch that the wavefront arrives at the outer boundary of our simulation domain. After $t>R_o$, waves start propagating out of the simulation domain. Lower Panel: relative errors between the numerical solutions and the exact analytic solution. In FEM, spatial resolution is the main driver of errors. By increasing the spatial resolution, the errors can be reduced significantly. \label{sphere_energy}}
\end{figure}

Figure~\ref{sphere_energy} compares the numerical results (stars) with the analytical ones (solid line) from Eq.~(\ref{totalenergy}). When the wavefront has not reached the outer boundary, namely, $R_i<t<R_o$, the numerical results agree with the analytic ones very well. After $t>R_o$ (vertical dashed line), the wavefront starts getting out of the simulation domain. The total energy inside the bulk of our simulation domain becomes steady (flat) as, in this case, the energy getting into the simulation domain equals the energy getting out of the domain.

\section{Gravitational Waves in a potential well} \label{sec:GWP}
In this section, we test the accuracy and robustness of our code in simulating GWs traveling in a potential well. We first introduce the basic equations and then discuss the numerical setups of solving these equations. 

\subsection{Basic equations}
We assume a metric $g_{\mu\nu}$ given by
\begin{equation}
ds^2=-(1+2\psi)dt^2+(1-2\psi)dx^2\,, \label{metric}
\end{equation}
and the gravitational waves as a perturbation to the background metric
\begin{equation}
	g_{\mu\nu}=g_{\mu\nu}^{0}+h_{\mu\nu}\quad,
\end{equation}
where $g_{\mu\nu}^{0}$ is the background metric. Following Ref.~\cite{Peters}, we neglect higher order non-linear terms and arrive at the propagation equation for gravitational waves $h_{ij}$
\begin{equation}
\nabla^2h_{ij}-(1-4\psi)\frac{\partial^2}{\partial t^2}h_{ij} = 0 \,.
\end{equation}

Using the eikonal approximation by Ref.~(\cite{Baraldo:1999ny}) (also see Ref.~\cite{GR_steven_weinberg}), the gravitational wave tensor can be represented as 
\begin{equation}
h_{ij}=h e_{ij}\,,
\end{equation}
where $e_{ij}$ is the polarization tensor. In this work, we assume that the polarization tensor does not change during the propagation of GWs. Then we obtain the so-called scalar wave equation
\begin{equation}
c^2\nabla^2h-\frac{\partial^2h}{\partial t^2} = 0 \,, \label{Wave_potential}
\end{equation}
where $c^2=1/(1-4\psi)$ is the speed of wave. 

Unlike the conventional wave equation, one notable feature of this equation is that the wave speed is not a constant but is dependent on the potential well. The varying wave speed plays an important role in the propagation of GWs in a potential well. Moreover, given the fact that $\psi<0$, we have $c<1$. The presence of potential thus delays the propagation of GWs. This phenomenon is known as the Shapiro time-delay~\cite{Shapiro}. 

\subsection{Plane wave approximation at the simulation box}
\begin{figure}
{\includegraphics[width=\linewidth]{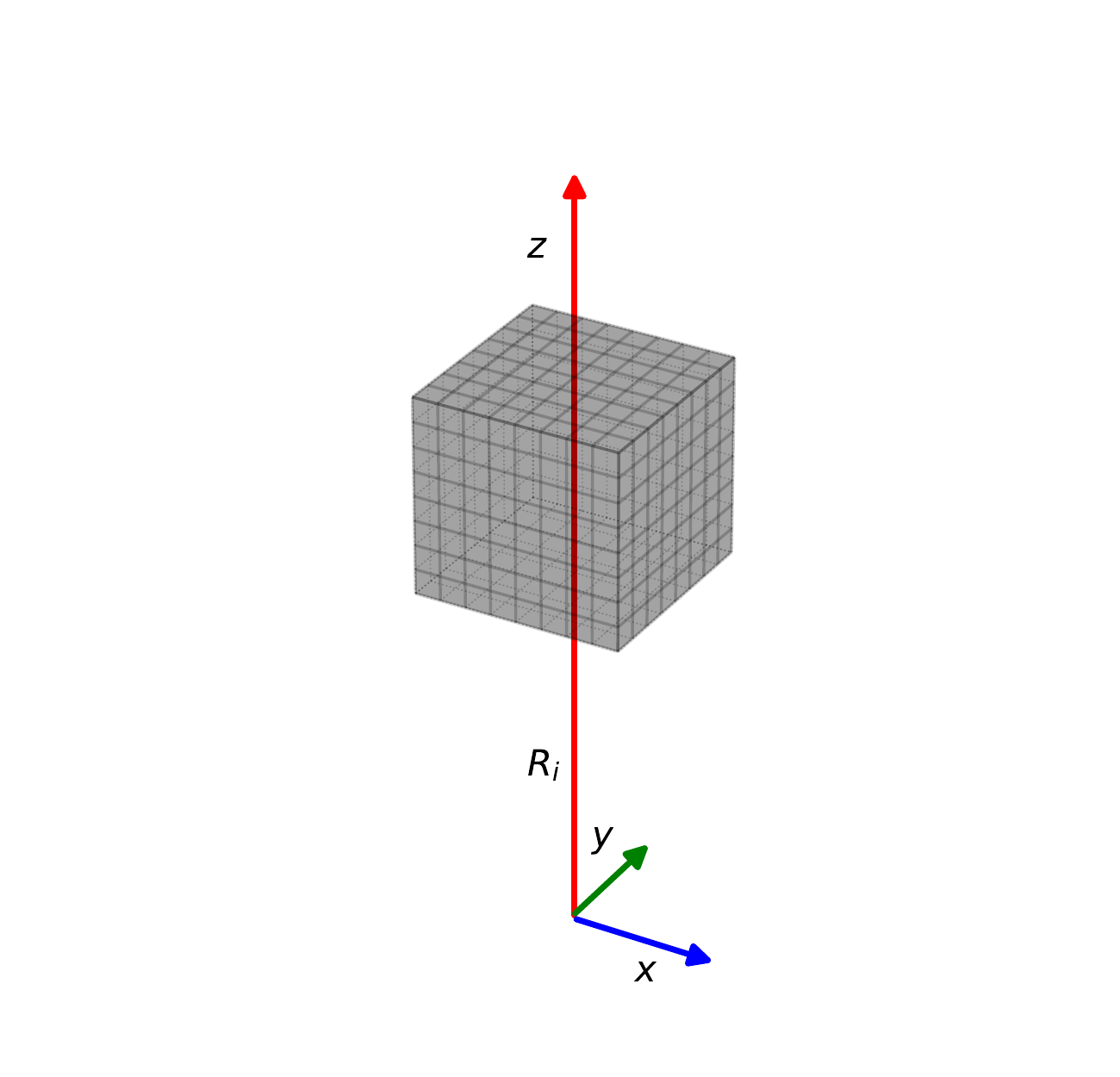}}
\caption{The domain of our simulations. $R_i$ is the distance from the origin of the simulation domain to the source of the GWs.  \label{Cubeboundary}}
\end{figure}

Figure~\ref{Cubeboundary} shows the domain and setups of our simulations, where $R_i$ is the distance from the source to the surface of the simulation domain. We assume that the source of GWs is far away from the simulation domain. The incident wave train travels along the $z$-axis and enters the surfaces of the simulation domain at $z=0$. Due to the long-range nature of gravity produced by the scatterer, the incident waves suffer the Shapiro time delay on their way from the source to the scatterer, which adds up a shift of phase to the wave function at the boundary of the simulation domain. Since the source is distant from the scatterer and the length of the simulation domain is small relative to $R_i$, the wave function at the simulation domain can be well approximated by plane waves with a constant phase shift $\Delta t = -2M\ln (2 R_i/d)$, where $d$ is the distance of the scatterer to the boundary of the simulation domain.  Moreover, we choose the size of the simulation domain large enough so that the scatterer does not distort the wave front initially at the boundary of the simulation domain. Such distortion effect can be accurately and robustly determined using the null geodesic equations, which are provided in appendix~\ref{Appdex:lightrays}. After these considerations, we set the boundary condition at $z=0$ as
\begin{align}
	\left \{
	\begin{aligned} 
h(t)|_{z=0}&= \frac{Q(t-R_i/c)}{4\pi R_i}\,\\
\dot{h}(t)|_{z=0}&= \frac{\dot{Q}(t-R_i/c)}{4\pi R_i}\,
\end{aligned}
\right. \,,
\end{align}
where $Q(t)$ is the waveform of the source and the dot denotes the derivative of time. 

In this subsection, we choose $Q$ as a sine function. For convenience, the initial phase of the source is so chosen that it cancels the change of phase due to the Shapiro time-delay when waves arrive at the simulation domain. As such, the boundary condition at $z=0$ can be represented as
\begin{align}
	\left \{
	\begin{aligned} 
h(t)|_{z=0}&= \sin(\omega t)\,\\
\dot{h}(t)|_{z=0}&= \omega \cos(\omega t)\,
\end{aligned}
\right. \,,
\end{align}
where $\omega=2\pi f$ and $f$ is the frequency of the wave. The amplitude of the incident wave is normalized as unity at $z=0$.  Moreover, we choose the zero point of the temporal axis at the epoch of the wavefront arriving at the surface of the simulation box. For other surfaces of the simulation domain, we adopt the absorbing boundary conditions as explained in Eq.~(\ref{absorbing}). The absorbing boundary conditions aim to guarantee that GWs $h$ do not reflect into the simulation box when they arrive at the boundaries. In addition to the boundary conditions, we set the initial condition inside the simulation domain as $$h = 0|_{t=0}\,,$$ which means that there are no waves in the simulation domain initially.

For $\psi$, we choose it as the potential generated by a solid spherical ball with uniform density
\begin{equation}
\psi=\left \{
	\begin{aligned} 
	&-\frac{M}{\left\|x-x_0\right\|} \quad &\left\|x-x_0\right\|> R_s\\
	&-M\frac{3R_s^2-\left\|x-x_0\right\|^2}{2R_s^3}\quad &\left\|x-x_0\right\|\le R_s
	\end{aligned}
	\right. \,,
\end{equation}
where $R_s = 2M$ is the Schwarzschild radius and $x_0$ is the center of the potential well. Note that, $\psi$ is continuous up to the first order of derivative, namely, $\psi\in C^1(\Omega)\subset H^1(\Omega)$. We choose the mass of the scatterer as
\begin{equation}
	M=10^5 M_{\odot}\,.
\end{equation}

\begin{figure*}
{\includegraphics[width=\linewidth]{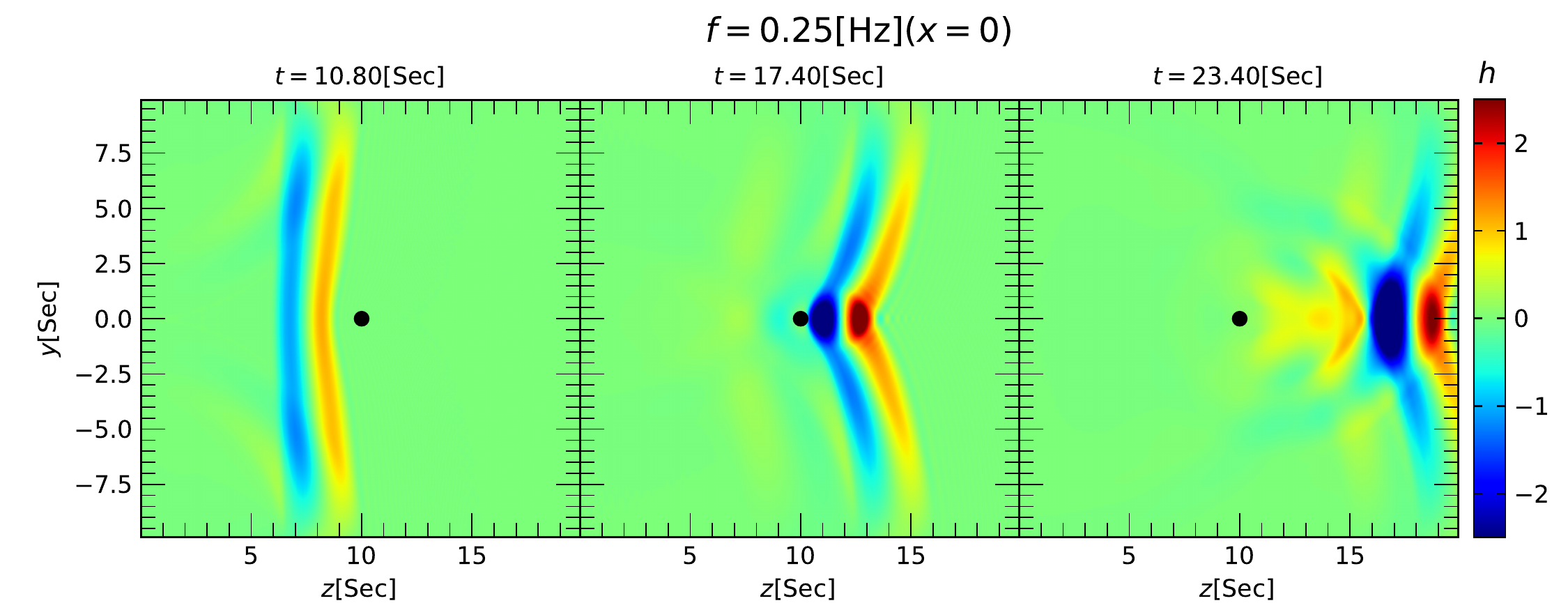}}
\caption{The waveform of a finite wave train in a potential well. From left to right, the panels show waveforms at different times. The waves enter the simulation domain at $z=0$ and then travels normally along the $z$-axis. The waveform is a sine function but only lasts for one period. The snapshots are taken along the $z-y$ plane at $x=0$. The color bar to the right shows the amplitude of waves $h$. The black dots indicate the position of the scatterer. As the wave speed is dependent on the potential well $c^2=1/(1-4\psi)$, waves travel faster far away from the center than those close to the center. The leading wavefront is bent toward the center (left panel). Unlike the thin-lens model as in the literature, after the scatterer, the deflection of the wavefront is so large that the wavefront can get warped, which leads to a complex 3D shape of the wave zone even when the waves get out of the potential well of the scatterer (right panel). 
\label{leadingwavefront}}
\end{figure*}

In this work, we first simulate the behavior of the leading wavefront for a finite wave train in a potential well, we choose the input wave train as a sine function but only lasting for one period
\begin{align}
h(t)|_{z=0} = \left \{
\begin{aligned} 
&\sin(\omega t)\,, & t \le \lambda \\
&0 \,,& t > \lambda
\end{aligned} 
\right. \,,
\end{align}
where $\lambda$ is the period. We choose the simulation domain as a cube. We choose the refinement of the simulation domain along each dimension in 3D space as $2^9$. The size of the simulation box is taken as $20\,[\rm sec]$ along one side.  We place the scatterer at the center of the box. As shown in appendix~\ref{Appdex:lightrays}, in this case the scatterer is distant from the boundary surface of the incident waves. As such, the distortion effect of the scatterer at boundaries is negligible. 

Figure~\ref{leadingwavefront} shows the waveform of the finite wave train in a potential well. We choose $\lambda=4[\rm sec]$ for illustrative purposes. From left to right, the panels show waveforms at different times. The snapshots are taken along the $z-y$ plane at $x=0$. The color bar to the right shows the amplitude of waves $h$. The black dot indicates the position of the scatterer. As the wave speed is dependent on the potential well $c^2=1/(1-4\psi)$, waves travel faster far away from the center than those closer to the center. The leading wavefront is bent toward the center (left panel). After the scatterer, the wavefront is warped, which leads to a complex 3D shape of the wave zones even when the waves get out of the potential well of the scatterer (right panel). 

\subsection{Numerical results against analytical predictions on the wavefront}
Unlike the rest part of waves, the leading wavefront represents the transport of energy from one place to another. Therefore, it is strictly subject to the constraint of causality. Moreover, the hypersurface of the wavefront (denoted as $S$) is a null hypersurface. Its normal vector $k^a=\nabla^a S$ is a null vector $k^ak_a=0$. As shown in \cite{wald2010general} (see, Eq.(4.2.37)), the integral curves of $k^a$ are null geodesics, which satisfy $k^a\nabla_a k^b=0$ if $k^a$ is presented in terms of an affine parameter. As such, the behavior of the leading wavefront can be traced by null geodesics, which provide an independent way to test our numerical results. As shown in the appendix, the equations for null geodesics can be presented as (see appendix~\ref{Appdex:lightrays} for details)
\begin{align}
\frac{d^2\mu}{d\varphi^2}+\mu=-\frac{1}{2}\frac{\partial}{\partial \mu}\left[\mu^2\zeta(M,\mu)\right]\,,\label{eqnongeo}
\end{align}
where $\mu=1/r$ and $r$ is the distance to the center of the mass. $\varphi$ is the azimuthal angle in the spherical coordinates. The definition of $\zeta(M,\mu)$ is provided in the appendix~\ref{Appdex:lightrays}. As the incident wave is a plane wave at $z=0$, the spacial components of its normal vector $k^a$ are parallel to the $z$-axis. Therefore, we set the initial conditions for Eq.~(\ref{eqnongeo}) at $z=0$ as
\begin{align}
    \mu(\varphi) \rightarrow \frac{1}{b}\sin \varphi\,,\label{LRboundary1}
\end{align}
where $b$ is the impact factor. Eq.~(\ref{LRboundary1}) simply describes a straight line in spherical coordinates along the $z$-axis, which is also the solution of Eq.~(\ref{eqnongeo}) in the absence of the scatterer $\zeta(M,\mu)= 0$.

\begin{figure*}
{\includegraphics[width=\linewidth]{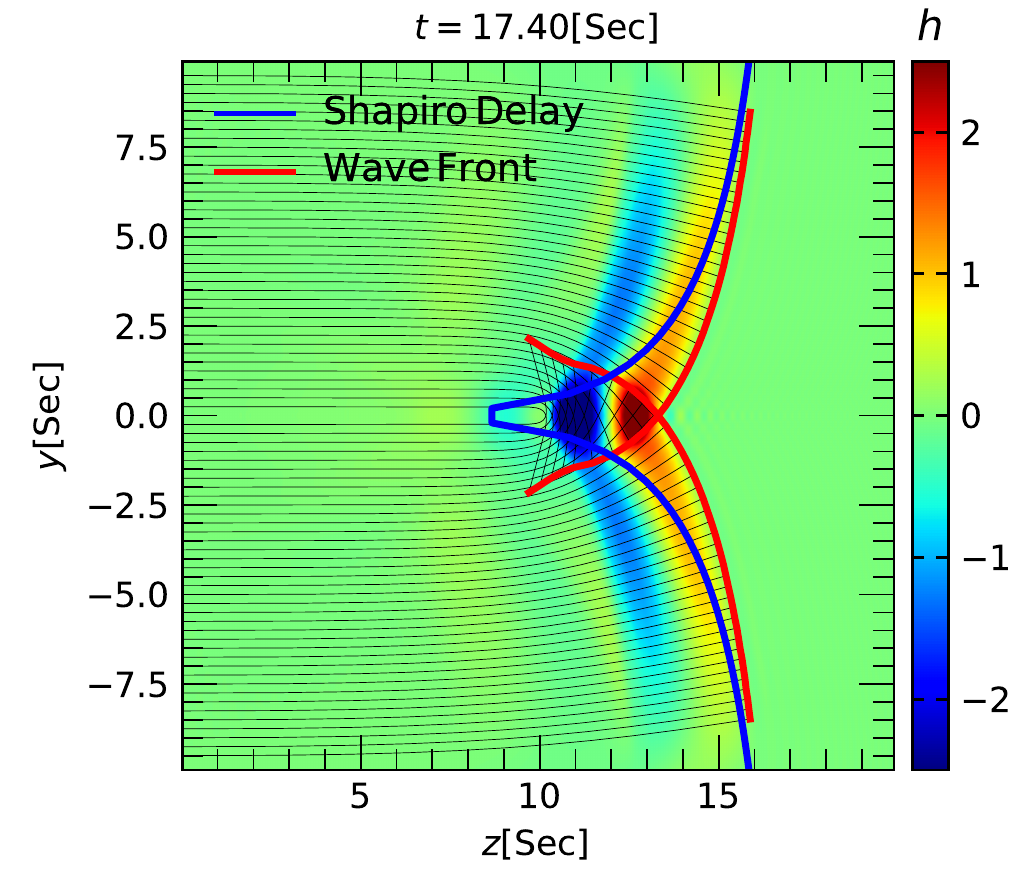}}
\caption{The integral curves of the normal vectors of the wavefront(the bundle of thin solid black lines). These curves follow the equation of null geodesics. The wavefront predicted by the null-geodesics (red solid line) agrees very well with simulations (colour shaded regions). Moreover, in the thin-lens regimes that are far away from the scatterer, our numerical results also agree with the predictions of the Shapiro-time delay (blue solid line) very well.\label{nullgeodesics}}
\end{figure*}

Figure~\ref{nullgeodesics} shows the trajectories of the null geodesics of the normal vectors $k^a$ of the wavefront obtained using Eq.~(\ref{eqnongeo}) (the bundle of thin solid black lines). The wavefront predicted by null geodesics at $t=17.40[{\rm sec}]$ (solid red lines) agrees very well with those of simulations (shaded regions). Using null geodesics, it is also clear that after the scatterer, the wavefront get warped and the signals of the leading wavefront on the $z$-axis does not come from the signals traveling on the $z$-axis itself, but from the wavefront circling around it. 

In addition to the tests using null geodesics, we also test our numerical results using the analytic expressions of the Shapiro-time delay. To to this, we introduce a new notion
\begin{align}
t_{\rm Shapiro}= \int_0^{t}c \,\mathrm{d}t\approx \int_0^{t}(1+2\psi)=t-\Delta t_{\rm Shapiro}\,,
\end{align}
where $c$ is the speed of wave $c=\sqrt{1/(1-4\psi)}$. For a point mass, the Shapiro time delay $\Delta t_{\rm Shapiro}$ is given by
\begin{align}
\Delta t_{\rm Shapiro} = -2\int \psi dz = 2M \int_{z_i}^{z_f}\frac{dz}{\sqrt{z^2+b^2}}\,, 
\end{align}
where $b$ is the impact factor and the integration is taken along the $z$ direction under the thin-lens assumption. ${z_i}$ and ${z_f}$ are the initial and final positions of the integration limits. In this work, we take $z_i=-10[{\rm sec}]$ and ${z_f}=7.40[{\rm sec}]$ which is the $z$ component of the position of the wavefront relative to the scatterer without delay at time $t=17.40[{\rm sec}]$. Fig.~\ref{nullgeodesics} shows that in regimes that are far away from the scatterer (where the thin-lens assumption is valid), our numerical results agree with the predictions of the Shapiro-time delay (blue solid line) very well.

\begin{figure*}
{\includegraphics[width=\linewidth]{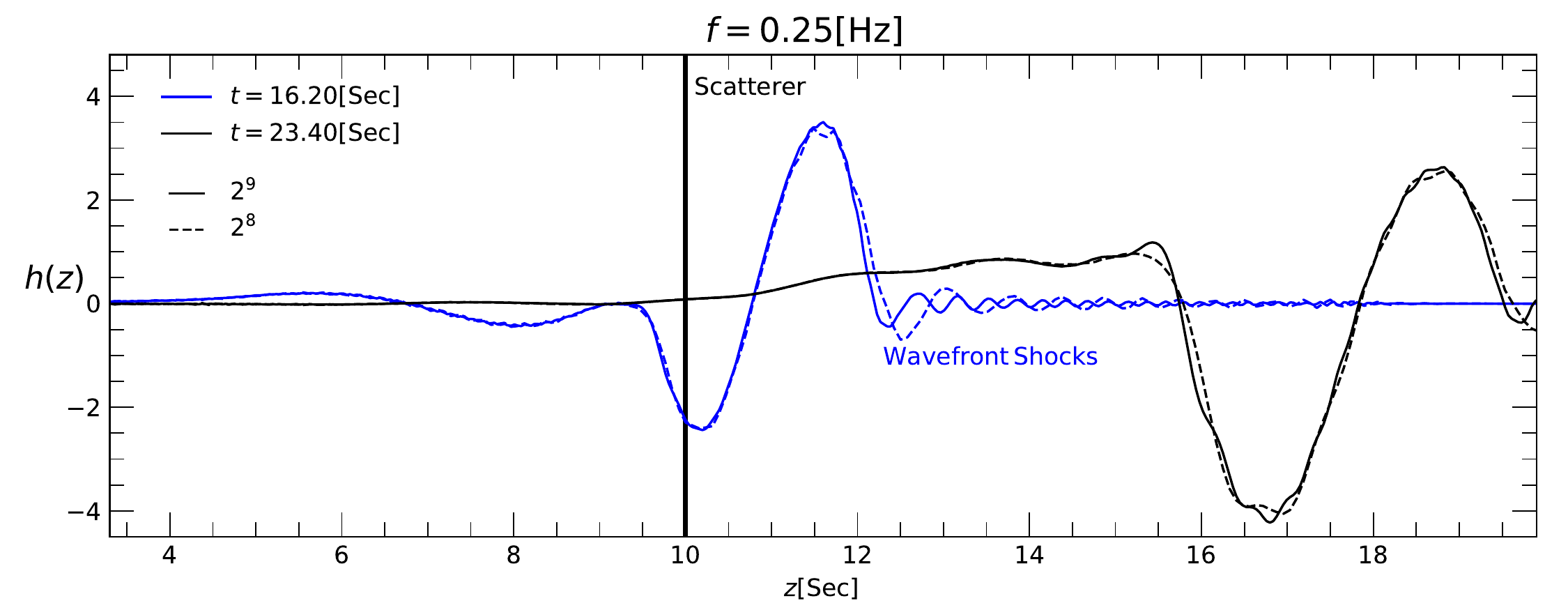}}
\caption{Numerical convergence tests for waveforms for simulations with different resolutions. The dashed and solid lines are for simulations with $2^8$ and $2^9$ refinements along each dimension in 3D space, respectively.
The solid black vertical line indicates the position of the scatterer. The overall waveforms for simulations with $2^8$ and $2^9$ refinements agree very well. However, a high-resolution run can suppress the wavefront shocks, which are numerical artifacts.\label{resolution}}
\end{figure*}

Moreover, we demonstrate the robustness of our numerical results. To do this, we compare the results obtained from the runs with refinements of $2^8$ and $2^9$ for wave functions at different times along the $z$-axis. From Fig.~\ref{resolution}, the waveforms for simulations with $2^8$ (dashed lines) agree with $2^9$ refinements (solid lines) very well, which means that even with $2^8$ refinement, it already yields good accuracy. However, a high-resolution simulation, such as $2^9$ refinement, can suppress the wavefront shocks, which are numerical artifacts appearing in the wave functions after the scatterer. 

\subsection{Numerical results against analytical predictions on waveform}
 
After testing the leading wavefront, we further test the waveform of our simulations against analytical predictions. In this new test, we take the incident wave function as a continuous sine function. Moreover, since the incident waves are of azimuthal symmetry, we choose the simulation domain as a cylinder instead of a cube to minimize the effects of boundaries. The incident waves are assumed to travel along the axis of the cylinder($z$ axis). The length of the cylinder is taken as $70[\rm sec]$ and the radius is taken as $25[\rm sec]$. To make it easier to be compared with analytical predictions, we simulate the waves with the evolution time long enough that the waves in the simulation domain can reach a steady state. In this case, the waveform can be analyzed in the Fourier space.  
\begin{align}
    h(\vec{x},t) = h(\vec{x})e^{-i\omega t}\,,
\end{align}
where $\omega = 2\pi f$ is the angular frequency. Equation~(\ref{Wave_potential}) then becomes
\begin{align}
    (\nabla^2+\omega^2)h=4\omega^2\psi h \label{wavefourier}\,.
\end{align}
Compared with Eq.~(\ref{Wave_potential}), the potential $\psi$ appears as a source term in Eq.~(\ref{wavefourier}). As such, the wave speed in Eq.~(\ref{wavefourier}) is unity and the wave number $\tilde{k}$ equals to the angular frequency $\tilde{k}=\omega$, which is different from that in Eq.~(\ref{Wave_potential}).

Outside the Schwarzschild radius of the scatterer, the potential $\psi$ of the scatterer can be described by $-M/r$. In this case, Eq.~(\ref{wavefourier}) can be solved analytically(e.g. ~\cite{Peters}). However, unlike Eq.~(\ref{Wave_potential}) which is hyperbolic, Eq.~(\ref{wavefourier}) is elliptic, which are specific to the ``boundary-value" problems~\cite{nla:cat-vn1414651}. The solutions of Eq.~(\ref{wavefourier}) are strongly dependent on the boundary conditions imposed. Since the wave functions are of azimuthal symmetry, we impose the following ansatz as boundaries 
\begin{align}
    \tilde{h}=e^{i\tilde{k}z}f(\xi)\,,
\end{align}
where $\xi=r-z$ and $r=\left\|x-x_0\right\|=\sqrt{x^2+y^2+z^2}$(the scatterer is placed at the origin). Since $f(\xi)$ does not depend on the direction of $\tilde{k}$, $\tilde{h}$ only has Fourier modes with wave vectors $\tilde{k}$ along the $z$-axis. As such, the above ansatz implicitly assumes that the wave vector $\tilde{k}$ of the outgoing waves at infinity is still along the $z$-axis. Under such assumption, the solution of Eq.~(\ref{wavefourier}) is given by
\begin{align}
\tilde{h} =Ce^{i\tilde{k}z}{_1F_1}(i\omega M,1,i\omega \xi)\,,
\end{align}
where ${_1F_1}$ is the confluent hypergeometric function and $C$ is a complex constant (detailed discussions are provided in Appendix~\ref{Appdex:Analytical}). When $M=0$
\begin{align*}
 {_1F_1}(0,1,i\omega \xi) = 1\,.   
\end{align*}
Wave function $\tilde{h}$ simply goes back to the plane wave. Moreover, it is worth noting that $\tilde{h}$ can not describe waveform on the $z$-axis because ${_1F_1}$ is not well defined when $\xi=0$.

If $r\gg R_s$ and $\xi\gg 1$, the confluent hypergeometric ${_1F_1}(i\omega M,1,i\omega \xi)$ can be expanded around $i\omega \xi=\infty$ as
\begin{align}
{_1F_1}(i\omega M,1,i\omega \xi)\approx \xi^{-i\omega M}\left[1+\frac{\omega^2M^2}{\xi} \right]\,.
\end{align}
In this case, the wave function can be written as
\begin{align}
\tilde{h}(x,y,z) \approx  Ce^{i\tilde{k}z} (\sqrt{x^2+y^2+z^2}-z)^{-i\omega M}\left[1+\frac{\omega^2M^2}{\sqrt{x^2+y^2+z^2}-z} \right]\,.
\end{align}
Note that $\tilde{h}(x,y,z)$ are complex values. The spatial waveform can then be obtained by taking the real part of $\tilde{h}(x,y,z)$. The complex constant $C$ is independent of $(x,y,z)$, which can be determined by matching the amplitude and phase of $\tilde{h}(x,y,z)$ with the incident waves. An advantage of the above equation is that it can be evaluated efficiently in the numerical processes. We therefore use it to test our simulation results.

\begin{figure*}
{\includegraphics[width=\linewidth]{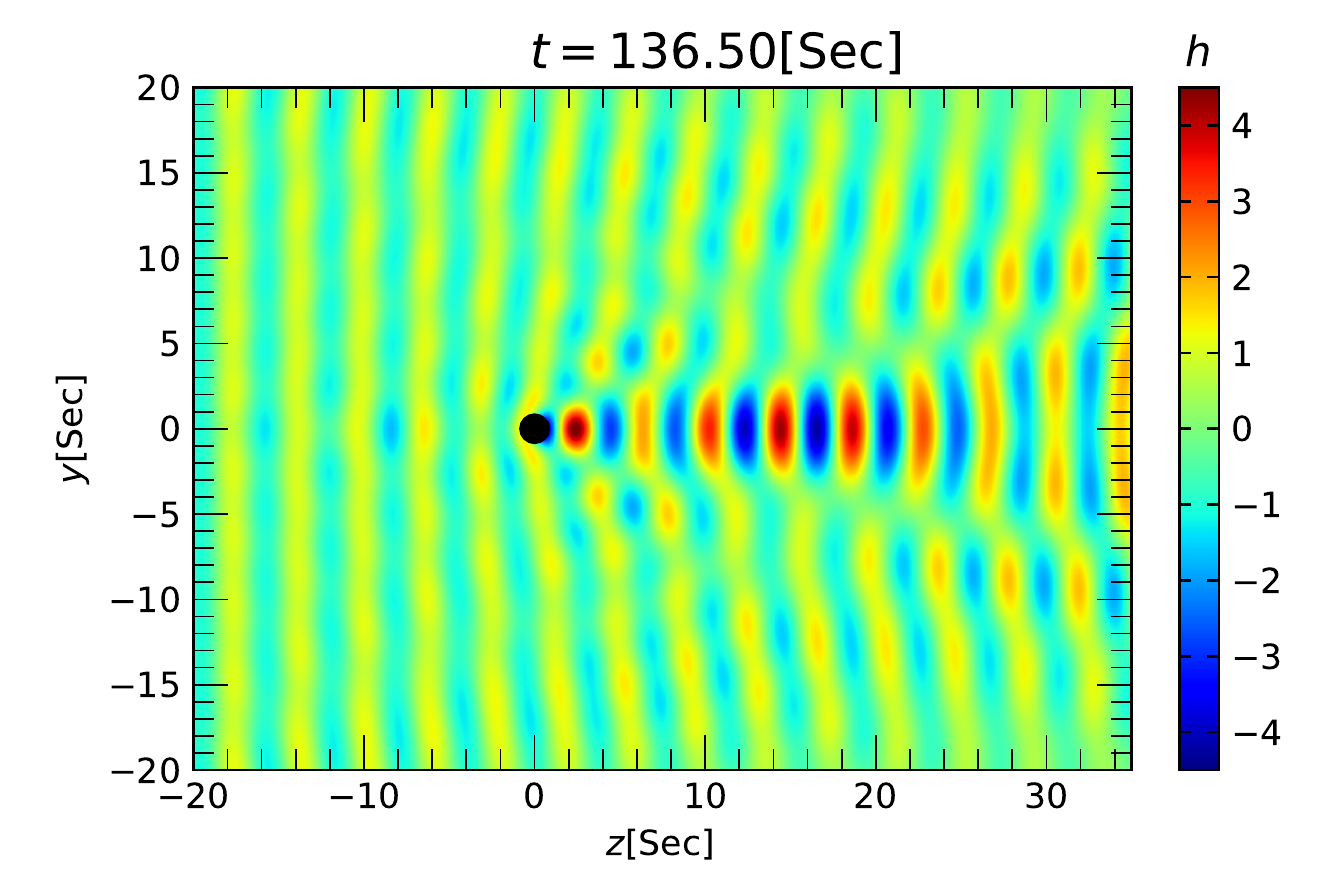}}
\caption{The waveform of a continuous wave train in a potential well. The snapshot is taken along the $z-y$ plane with $x=0$ at $t=136.50[\rm Sec]$. Since the incident waves are of azimuthal symmetry, in this case the simulation domain is taken as a cylinder with the incident waves traveling along the direction of the axis of the cylinder($z$-axis). The wavefrom of the incident waves is a continuous sine function with a frequency of $f=0.25\,[{\rm Hz}]$. The time evolution is taken long enough so that the waveform is in a steady state. The black dot indicates the position of the scatterer. The signals after the scatterer show a clear diffraction pattern with strong signals along the $z$-axis.  \label{2Dsteadystat}}
\end{figure*}

Figure~\ref{2Dsteadystat} shows a snapshot of our simulation along the $z-y$ plane at $x=0$. The incident wave travels along the $z$-axis with a waveform as a continuous sine function. The frequency of the incident wave is $f=0.25\,[{\rm Hz}]$. The snapshot is taken at $t=136.50[\rm Sec]$, for which the time evolution is long enough so that the waveform is in a steady state. We have checked that further increasing the evolution time does not change the waveform appreciably. The black dot indicates the position of the scatterer. The signals after the scatterer show a clear diffraction pattern with strong signals along the $z$-axis.

\begin{figure*}
{\includegraphics[width=\linewidth]{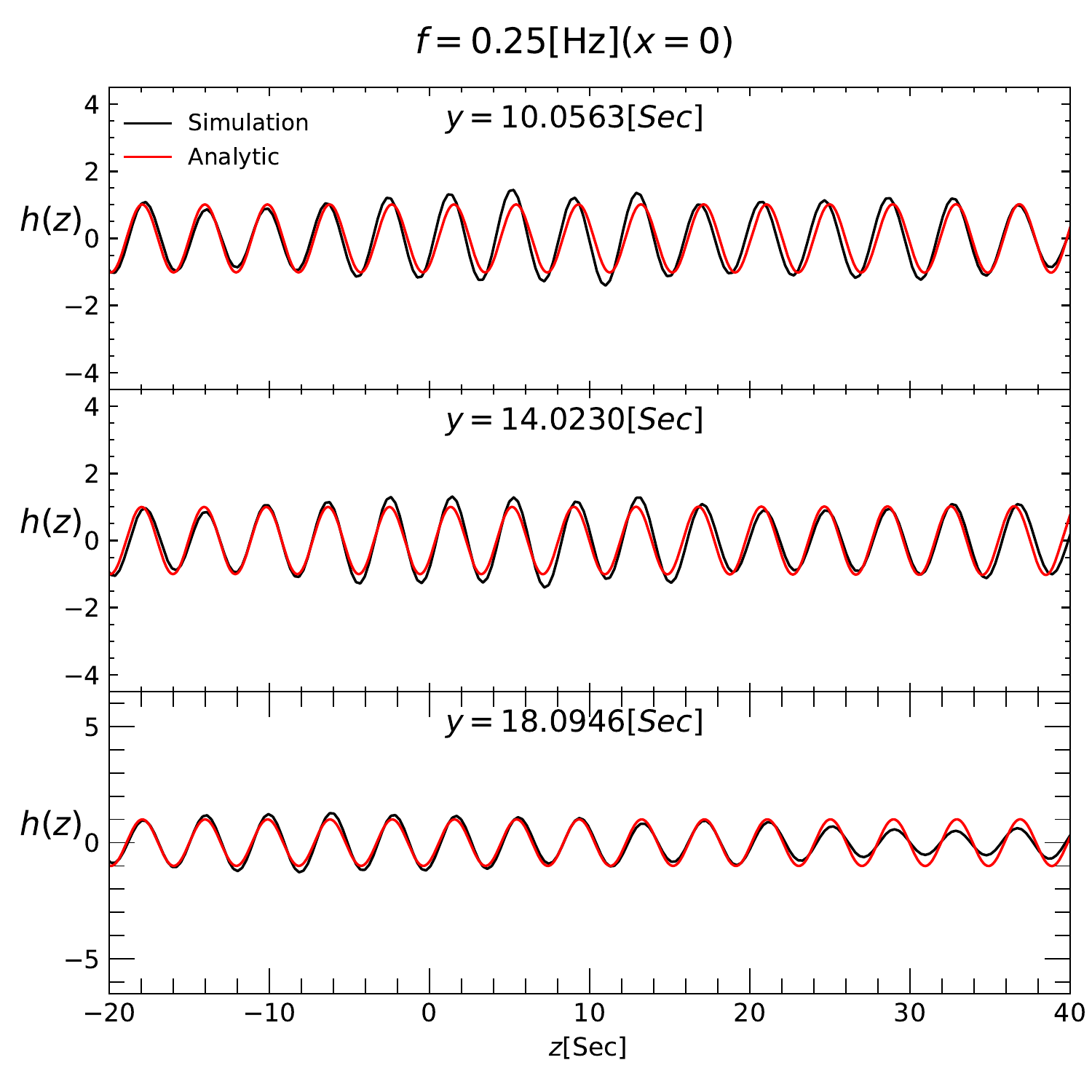}}
\caption{The 1D waveform of simulations against analytical predictions. The waveforms are taken along different lines that are parallel to the $z$-axis. The distances of these lines to the $z$-axis are taken as $y=10.0563[\rm Sec]$, $y=14.0230[\rm Sec]$ and $y=18.0946[\rm Sec]$, respectively. The black lines show the waveform from simulations and the red lines show the analytical predictions. The simulations agree with the analytical predictions very well. 
\label{1Dsteadystat}}
\end{figure*}

Figure~\ref{1Dsteadystat} compares the wavefroms of our simulations (black lines) with the analytical predictions (red lines). The wavefroms are taken along 1D lines that are parallel to the $z$-axis. Since the wave function and simulation domain are of azimuthal symmetry, we only extract these lines from the $z-y$ plane. The distances of these lines to the $z$-axis are chosen as $y=10.0563[\rm Sec]$, $y=14.0230[\rm Sec]$ and $y=18.0946[\rm Sec]$, respectively. The waveform of our simulations agree with the analytical predictions very well. 

\section{Conclusions}  \label{sec:cl}
The detection of gravitational waves in the low-frequency regime provides a powerful tool to probe the properties of supermassive back holes (SMBHs) throughout cosmic history, which has profound implications in gravitational physics and galaxy formation physics. However, due to the wave nature of GWs, wave effects such as diffraction or interference may hamper the future GW experiments to accurately infer the source information from the observed signals. Accurately pinning down such effects, therefore, is urgently called for.

The wave effects of GWs have already been discussed in the literature. However, the previous analyses focus on the frequency-domain, coupled to some implicit assumptions, such as that GWs are in a steady-state (e.g. Ref.~\cite{Peters}) or stationary state (e.g. Ref.~\cite{Suyama:2005mx}). These analyses also assume that the wave zone of GWs is infinite, which neglects some important wave effects such as the causality and wavefront effects. For a time-finite GW signal, these two effects may have significant impacts on the waveforms. Contrary to common intuition that the time-domain and frequency-domain are equivalent, the wavefront effects in a potential well should be studied in the time-domain. This is because the propagation of wavefronts is a ``Cauchy" problem for hyperbolic equations, which is defined in part of the spacetime and is fundamentally local (more discussions can be found in section 10.1 of \cite{wald2010general}). If taking the Fourier transforms, the wave equations in the frequency-domain become elliptic (e.g. Helmholtz equations), which are specific to the ``boundary-value" problems~\cite{nla:cat-vn1414651}. The wave functions, in this case, are no longer determined by initial conditions but by boundary conditions. In this case, the boundaries in the future at infinity have to be specified, which can only be consistently set for simple cases, such as waves traveling in free space. However, if for a complicated problem, such as waves traveling in a potential well, the wave-zones may have complex shapes in 3D space and it is difficult to set consistent boundary conditions. In these cases, GWs should be studied in the time-domain directly.  

Motivated by these facts, in this paper, based on the publicly available code {\bf deal.ii} as well as stand-alone linear algebra libraries, such as PETSc, we have developed a code called {\it GWsim} to simulate the propagation of GWs in a potential well. We have tested our code using a monochromatic spherical wave from a point source. We find that not only for the waveforms at an individual observer but also for the conservation of energy, our numerical results agree with the analytical predictions very well (see Fig.~\ref{sphere_energy}). 

After testing our code, we have studied the propagation of wavefront in a potential well. We focus particularly on a time-domain specific problem, namely, the leading wavefront in a potential well. Since the integral curves of the normal vectors of the leading wavefront are null-geodesics, the leading wavefront can be traced by the null geodesics, which provide an independent way to test our numerical results. We find that the wavefront predicted by null geodesics agree with our numerical results very well (see Fig.~\ref{nullgeodesics}). In addition to the null geodesics, we also compared our numerical results with the analytical expressions of the Shapiro-time delay
\begin{align}
\Delta t_{\rm Shapiro} = -2\int \psi dz = 2M \int_{z_i}^{z_f}\frac{dz}{\sqrt{z^2+b^2}}\,.\label{shapiroformular}
\end{align}
We find that in the thin-lens regimes, namely, far away from the scatterer, our numerical results agree with the predictions of Eq.~(\ref{shapiroformular}) very well. However, near the optical axis, we find that Eq.~(\ref{shapiroformular}) can not give accurate predictions of the positions of the leading wavefront. The actual wavefront travels faster than the predictions of Eq.~(\ref{shapiroformular}). These results suggest that the analyses presented in~\cite{Suyama:2020lbf} and \cite{Ezquiaga:2020spg} is only valid in the thin-lens regimes since the authors there use Eq.~(\ref{shapiroformular}) to predict the speed of GWs. Moreover, it is also worth noting that unlike the conventional images in geometric optics, GWs can not be sheltered by the scatterer due to wave effects. The signals of GWs can circle around the scatterer and travel along the optic axis until arrive at a distant observer(see e.g. Fig.~\ref{nullgeodesics}), which is an important observational consequence of GWs in such a system. 
 
In addition to the above conclusions, due to the linearity of the wave equation and the geometric unit, our numerical results can serve as templates. The mass of the scatterer used in this work can be rescaled to other values that are of astrophysical interest. For instance, if we want to obtain the results for a scatterer with a mass of $M=10^7 M_{\odot}$, we only need to re-scale the temporal and spatial axes of our numerical results by $100$ times. 

Finally, this work is the first attempt to simulate the wave effects of GWs in the time domain. Therefore, following the previous work in the literature~\cite{ PhysRevD.34.1708,Deguchi,Schneider,Ruffa_1999,DePaolis:2002tw,Takahashi:2003ix,1999PThPS.133..137N,Suyama:2005mx,Christian:2018vsi,Zakharov_2002,Liao:2019aqq,Macquart:2004sh,PhysRevLett.80.1138,Dai:2018enj,PhysRevD.90.062003,Yoo:2013cia,Nambu:2019sqn}, we treat the potential well of the scatterer as Newtonian gravity, which is reasonable in this work, since we only focus on the leading wavefront, for which most signals come from regimes that are far away from the scatterer. However, to fully explore the effects of general relativity on GWs requires new development of the current code, which is beyond the scope of this paper. In a series of follow-up papers, we will extend our work to the spacetime of a realistic black hole, such as a Kerr black hole~\cite{Baraldo:1999ny}. The method presented in this paper will be a powerful tool for exploring GWs in a non-static potential well, which is of significant astrophysical interests. Moreover, we will also explore the effectiveness of using the propagation of GWs as a tool to test the theory of gravity~\cite{Chesler:2017khz,Belgacem:2019pkk,Koyama:2020vfc}. 

\section*{Acknowledgements}

J.H.H. acknowledges support of Nanjing University. This work is supported by the National Natural Science Foundation of China (12075116). The numerical calculations in this paper have been done on the computing facilities in the High Performance Computing Center (HPCC) of Nanjing University.

\section*{Data Availability}
The data and code underlying this article will be shared on reasonable request to the corresponding author.




\bibliographystyle{mnras}
\bibliography{myref.bib}




\appendix
\section{Spherical Wave in free-space} \label{Appdex:sw}
We adopt the following convention for the Fourier transform 
\begin{align}  
u(r,\theta,\phi,t)&=\int  u(r,\theta,\phi,\omega)e^{-i\omega t}\, \mathrm{d}t \,,\nonumber\\
u(r,\theta,\phi,\omega)&=\frac{1}{2\pi}\int  u(r,\theta,\phi,t)e^{i\omega t}\, \mathrm{d}t \,.\nonumber
\end{align}
In the frequency-domain, Eq.~(\ref{strong}) in free-space becomes
\begin{equation}
\left[\nabla^2+(\omega/c)^2\right] u = 0\,.
\end{equation}
For a diverging spherical wave radiating from the origin, the general solution of the above equation has the form
\begin{equation}
	u(r,\theta,\phi,\omega) =\sum_{lm}a_{lm}(\omega)h_l^{(1)}\left(\frac{\omega}{c}r\right)Y_{lm}(\theta,\phi)\,,\label{u_general}
\end{equation}
where $h_l^{(1)}(x)$ is the spherical Hankel functions of the first kind, which is the solution of the spherical Bessel equation satisfying the radiation boundary condition at infinity. $Y_{lm}(\theta,\phi)$ is the spherical harmonics. $a_{lm}(\omega)$ are unknown coefficients, which can be determined by the boundary conditions.

We assume that the boundary condition for a spherical wave is given at the surface of a sphere with a radius of $R$. $$u(r,\theta,\phi,\omega)|_{\partial \Omega}=f_R(\theta,\phi,t)\,,$$ where $f_R(\theta,\phi,t)$ is a known-function. In the frequency-domain, we have 
\begin{equation}  
f_R(\theta,\phi,\omega)=\frac{1}{2\pi}\int f_R(\theta,\phi,t)e^{i\omega t}\, \mathrm{d}t\,.
\end{equation}
We expand $f_R(\theta,\phi,\omega)$ in terms of spherical harmonics
\begin{equation}  
f_R(\theta,\phi,\omega)=\sum_{lm}b_{lm}(w)Y_{lm}(\theta,\phi)\,.
\end{equation}

Comparing the above equation with Eq.~(\ref{u_general}) at the boundary, we obtain
\begin{equation}
	a_{lm}(\omega) h_l^{(1)}\left(\frac{\omega}{c}R\right) = b_{lm}(\omega)\,.
\end{equation}
Thus, we have
\begin{align}
	\begin{split}
u(r,\theta,\phi,t)&=\int  e^{-i\omega t} u(r,\theta,\phi,\omega) \, \mathrm{d}t \\
&=\int e^{-i\omega t} \sum_{lm} \frac{h_l^{(1)}(\frac{\omega}{c}r)}{h_l^{(1)}(\frac{\omega}{c}R)} b_{lm}(w)Y_{lm}(\theta,\phi)\, \mathrm{d}t \,. 
    \end{split}
\label{sphere_general}
\end{align}
The above equation demonstrates the Huygens-Fresnel principle, namely, the propagation of waves is completely determined by the surface of the wavefront. 

If the wavefront at the boundary is of spherical symmetry and is the same as the wavefront propagating there from the point source at the origin
\begin{equation}
f_R(\theta,\phi,t) = \frac{Q(t-\frac{R}{c})}{4\pi R}\,,
\end{equation}
there will be only monopole $l=0$ left in the expansions of Eq.~(\ref{sphere_general}). Further noting that
\begin{equation}
h_0^{(1)}(x)= -\frac{i}{x}e^{ix}\,,
\end{equation}
Eq.~(\ref{sphere_general}) then reduces to
\begin{align}
	\begin{split}
u(r,\theta,\phi,t)&=\frac{R}{r}\int e^{-i\omega (t-\frac{r-R}{c})}f_R(\theta,\phi,\omega) \, \mathrm{d}t \\ &=\frac{R}{r}f_R(\theta,\phi,t-\frac{r-R}{c}) \\
&=\frac{Q(t-\frac{r}{c})}{4\pi r}
    \end{split}\,,
\end{align}
which is the same as the wave propagating directly to the radius $r$ from the origin.

In general cases, if $x=\frac{\omega}{c}r\gg1$
\begin{equation}
	h_l^{(1)}(x)\sim \frac{1}{x}e^{ix}(-i)^{l+1} \nonumber\,,
\end{equation}
we have
\begin{align}
	\frac{h_l^{(1)}(\frac{\omega}{c}r)}{h_l^{(1)}(\frac{\omega}{c}R)}\sim \frac{R}{r}e^{i\frac{\omega}{c}(r-R)}\,.
\end{align}
The wave function reduces to
\begin{align}
	\begin{split}
u(r,\theta,\phi,t)&=\frac{R}{r}\int e^{-i\omega (t-\frac{r-R}{c})}f_R(\theta,\phi,\omega) \, \mathrm{d}t \\ &=\frac{R}{r}f_R(\theta,\phi,t-\frac{r-R}{c})
    \end{split} \,.
\end{align}
This indicates that the anisotropic feature in the wave function at the radius $R$ can be preserved to a distant observer during the propagation of waves, if the waves travel in a free-space. The amplitude of the wave decreases by a factor of $\frac{R}{r}$, which is the same as the isotropic wave, which is due to the conservation of energy of waves.

\section{Null geodesics in curved spacetime} \label{Appdex:lightrays}
\begin{figure}
{\includegraphics[width=\linewidth]{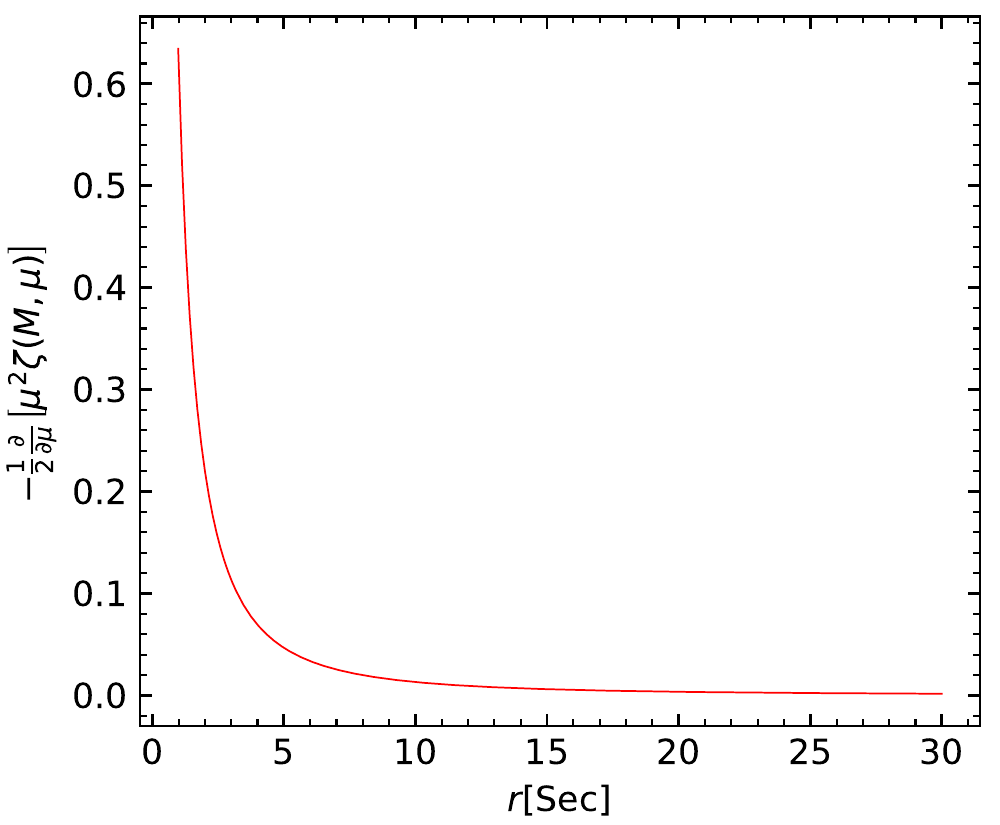}}
\caption{The right-hand side of Eq.(~\ref{muphi}) as a function of radius $r$. For $r\gg 2M$, $\frac{\partial}{\partial \mu}\left[\mu^2\zeta(M,\mu)\right]$ decays as $1/r^2$, which is much faster than that of the potential.
\label{zeta_r}}
\end{figure}

\begin{figure*}
{\includegraphics[width=\linewidth]{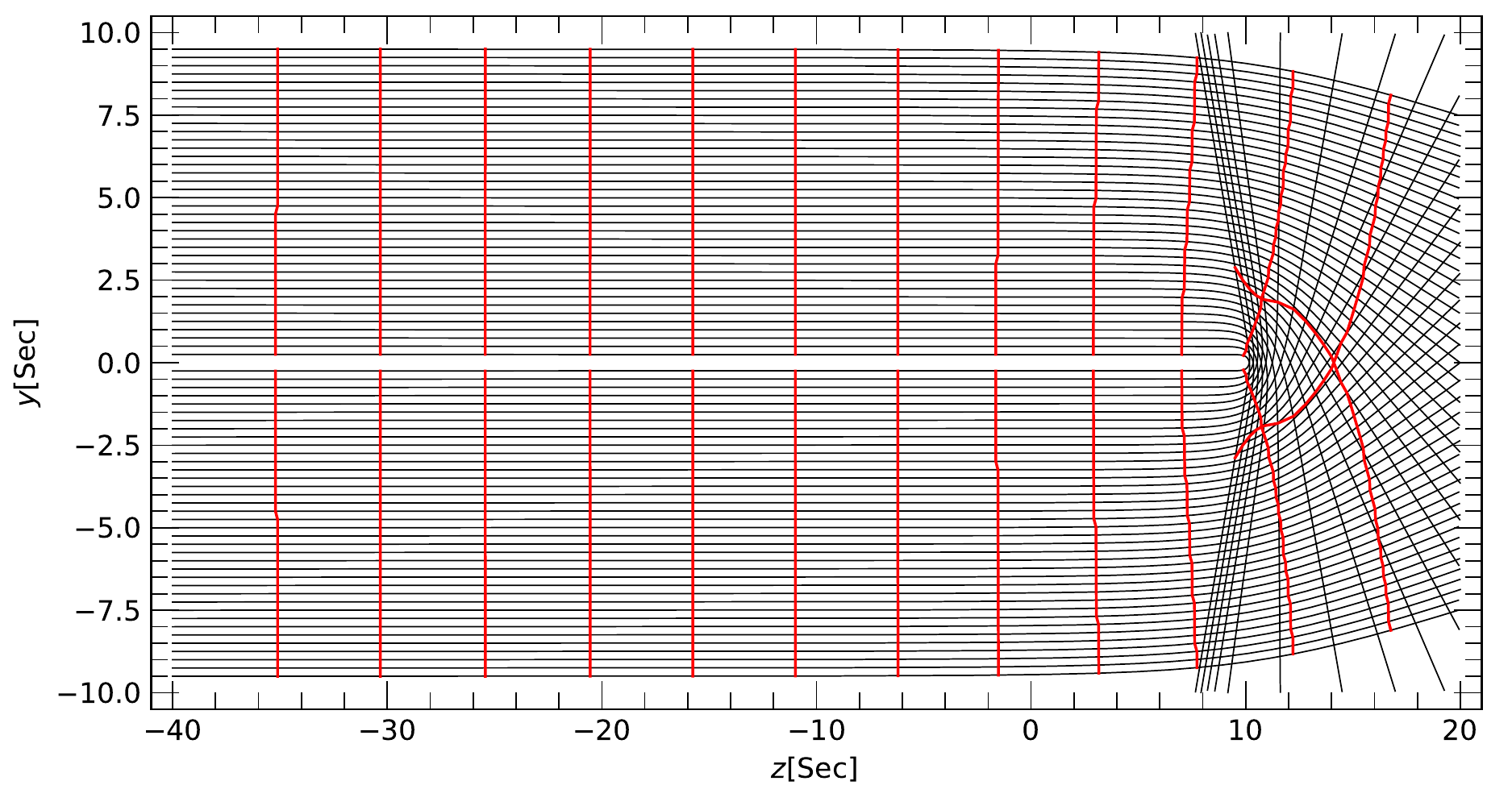}}
\caption{The trajectories of null geodesics (black solid lines) in the spacetime of the scatterer. The center of the scatterer is at $z=10[\rm sec]$. The initial conditions of Eq.~(\ref{LRboundary}) for the null geodecis equations are set at $z=-40 [\rm sec]$, $50 [\rm sec]$ away from the scatterer. The red curves indicate the wavefront at different killing times $t$. As Eq.~(\ref{muphi}) can not be applied to the $z$-axis, we do not show light rays on the $z$-axis. The impact of the scatterer on wavefront is only significant  when the wavefront is close to the scatterer within a radius of $10[\rm sec]$. However, outside this region, the distortion effect on the wavefront becomes negligible.
\label{zeta_wavefront}}
\end{figure*}

In this subsection, we describe the null geodesics in the spacetime of a point mass. We assume that the line element has a generic form
\begin{align}
ds^2 = g_{00}(M,r)dt^2+g_{11}(M,r)dr^2+r^2(d\theta^2+\sin\theta^2d\varphi^2)\,,
\end{align}
where $g_{00}(M,r)$ and $g_{11}(M,r)$ are functions of mass and radius from the point mass. For a null-geodesic, $\left(\frac{\partial}{\partial \tau}\right)^a$ denotes the tangent vector and $\tau$ is the affine parameter.
We assume that the initial position and the tangent vector of the geodesic lie in the ``equatorial plane" $\theta = \pi/2$ (this is always possible by choosing proper coordinates). The null condition $g_{ab}\left(\frac{\partial}{\partial \tau}\right)^a \left(\frac{\partial}{\partial \tau}\right)^b = 0$ gives
\begin{align}
g_{00}(M,r)\left(\frac{dt}{d\tau}\right)^2+g_{11}(M,r)\left(\frac{dr}{d\tau}\right)^2+r^2\left(\frac{d\varphi}{d\tau}\right)^2=0\,.\label{Null_condition}
\end{align}
As $\left(\frac{\partial}{\partial t}\right)^a$ is a static time-like killing vector, the quantity 
\begin{align}
E = -g_{ab}\left(\frac{\partial}{\partial \tau}\right)^a\left(\frac{\partial}{\partial t}\right)^b=-g_{00}\left(\frac{dt}{d\tau}\right)\,
\end{align}
along the geodesic $\left(\frac{\partial}{\partial \tau}\right)^a$ is a constant. Likewise, for the rotational Killing vector $(\frac{\partial}{\partial \varphi})^a$
\begin{align}
L = r^2\left(\frac{d\varphi}{d\tau}\right)\,
\end{align}
is a constant. Inserting $E$ and $L$ back to Eq.~(\ref{Null_condition}), we obtain
\begin{align}
g_{00}^{-1}(M,r)E^2+g_{11}(M,r)\left(\frac{dr}{d\tau}\right)^2+\frac{L^2}{r^2}=0\,.
\end{align}
Further by noting that
\begin{align}
\frac{dr}{d\tau}=\left(\frac{dr}{d\varphi}\right)\frac{L}{r^2}\,,
\end{align}
and using the condition
\begin{align}
g_{00}(M,r)g_{11}(M,r)=-1\,,
\end{align}
we arrive at
\begin{align}
\left(\frac{dr}{d\varphi}\right)^2-\frac{E^2r^4}{L^2}+\frac{r^2}{g_{11}}=0\,.\label{phimu2}
\end{align}
If we denote $\mu = 1/r$, the above equation can be cast into
\begin{align}
\left(\frac{d\mu}{d\varphi}\right)^2=\frac{E^2}{L^2}-\frac{\mu^2}{g_{11}}\,.\label{muphi0}
\end{align}
It is more convenient to write $g_{11}^{-1}(M,r)$ as
\begin{align}
	g_{11}^{-1}(M,r)=1+\zeta(M,\mu)\,.
\end{align}
Taking the derivative of Eq.~(\ref{muphi0}) and noting that $E$ and $L$ are constants, we obtain
\begin{align}
\frac{d^2\mu}{d\varphi^2}+\mu=-\frac{1}{2}\frac{\partial}{\partial \mu}\left[\mu^2\zeta(M,\mu)\right]\,. \label{muphi}
\end{align}
The above equation gives the trajectories of null geodesics in terms of $\mu$ and $\varphi$. However, to solve Eq.~(\ref{muphi}), we need to set the initial conditions. We assume that the source of the incident light rays are far away from the point mass and are parallel to the $z$-axis. In this case, $\zeta(M,\mu)\rightarrow 0$ and Eq.~(\ref{muphi}) has the solution
\begin{align}
    \mu(\varphi) \rightarrow \frac{1}{b}\sin \varphi\,,\label{LRboundary}
\end{align}
where $b$ is the impact factor. The above equation is simply a straight line in the spherical coordinates. We use Eq.~(\ref{LRboundary}) as initial conditions at the boundary of the simulation box. Comparing to Eq.~(\ref{metric}), we find
\begin{align}
    \zeta(M,\mu) = \frac{2\psi(M,\mu)}{1-2\psi(M,\mu)}\,,
\end{align}
and
\begin{align}
    \frac{\partial}{\partial \mu}\left[\mu^2\zeta(M,\mu)\right] = -\frac{8\mu^2M(M\mu+3/4)}{(2M\mu+1)^2}\,.\label{diff_zeta_mu}
\end{align}
Then Eq.~(\ref{muphi}) can be numerically solved for the rest trajectories of null geodesics. After getting the trajectories, the lengths of trajectories can be obtained by integrating the line element
\begin{align}
dl=\frac{1}{\mu}\sqrt{\frac{1}{\mu^2}\left(\frac{d\mu}{d\varphi}\right)^2+1}d\varphi\,.
\end{align}
Then the asymptotic time of the light rays can be further obtained by
\begin{align}
	t=\int \frac{dl}{c}\,,
\end{align}
where $c$ is the wave speed observed by asymptotic observer $c=dl/dt=\sqrt{1/(1-4\psi)}$. 

However,  in practice, Eq.~(\ref{LRboundary}) has to be set at a finite radius, where Eq.~(\ref{diff_zeta_mu}) should be small enough and its impact on the incident rays can be neglected. For this purposes, in Figure~\ref{zeta_r} we show the right-hand side of Eq.(~\ref{muphi}) as a function of radius $r$. In this calculation, we take $M=10^5M_{\odot}$, the same as in the main text. For $r\gg 2M$, $\frac{\partial}{\partial \mu}\left[\mu^2\zeta(M,\mu)\right]$ decays as $1/r^2$, which is faster than that of the potential. As a result, when $r > 10 [\rm sec]$, $\frac{\partial}{\partial \mu}\left[\mu^2\zeta(M,\mu)\right]$ becomes negligible.

Figure~\ref{zeta_wavefront} shows the trajectories of null geodesics in the spacetime of the scatterer. The initial conditions of Eq.~(\ref{LRboundary}) are set at $z=-40 [\rm sec]$, $50 [\rm sec]$ away from the scatterer. The red curves indicate the wavefront at different killing times $t$. Since Eq.~(\ref{muphi}) can not be applied to the $z$-axis, we do not show light rays on the $z$-axis. From Figure~\ref{zeta_wavefront}, the impact of the scatterer on wavefront is only significant when the wavefront is close to the scatter within a radius of $10[\rm sec]$. Outside this region, the distortion effect on the wavefront is negligible.

\section{Analytical solutions for wave equations}\label{Appdex:Analytical}

We work with the paraboloidal coordinates
\begin{equation}
    \left \{
    \begin{aligned}
    \xi &= r-z\,,\\
    \eta &= r+z\,.
    \end{aligned}
    \right.
\end{equation}
For $\psi=-M/r$, Eq.~(\ref{wavefourier}) can be cast into
\begin{align}
    \frac{4}{\xi+\eta}\left[ \frac{\partial}{\partial \xi}\left(\xi\frac{\partial h}{\partial \xi}\right)+\frac{\partial}{\partial \eta}\left(\eta\frac{\partial h}{\partial \eta}\right)+\frac{1}{\xi\eta}\frac{\partial^2h}{\partial \phi^2}\right] +\omega^2h +\frac{4\omega^2M}{\xi +\eta}h=0 \,. \label{paraboloidalwave}
\end{align}
Since the incident waves are of azimuthal symmetry, $h$ does not depend on $\phi$ and is only a function of $\eta$ and $\xi$. We, theorfore, choose the following ansatz
\begin{align}
    \tilde{h}=e^{i\tilde{k}z}f(\xi)=e^{i\tilde{k}\frac{\eta-\xi}{2}}f(\xi)\,.
\end{align}
Since $f(\xi)$ does not depend on the direction of $\tilde{k}$, $\tilde{h}$ only has Fourier modes with wave vectors $\tilde{k}$ along the $z$-axis. As such, the above ansatz implicitly assumes that the wave vector $\tilde{k}$ of the outgoing waves at infinity is still along the $z$-axis. Inserting the above expression into Eq.~(\ref{paraboloidalwave}), we obtain an equation for $f$
\begin{align}
\xi \frac{d^2}{d\xi^2}f+(1-i\tilde{k}\xi)\frac{d}{d\xi}f+M\omega^2f=0\,.
\end{align}
The above equation has solutions
\begin{align}
f =C{_1F_1}(i\omega M,1,i\omega \xi),\quad for \quad \xi \neq 0 \,,
\end{align}
where ${_1F_1}$ is known as the confluent hypergeometric function and $C$ is a constant, which can be determined by boundary conditions.  When $M=0$
\begin{align*}
 {_1F_1}(0,1,i\omega \xi) = 1\,.   
\end{align*}
Wave function $\tilde{h}$ simply goes back to the plane wave. Moreover, it is worth noting that ${_1F_1}$ is not well defined on the $z$-axis ($\xi=0$).

If we focus on regions that far away from the $z$-axis where $\xi\gg 1$, the confluent hypergeometric  ${_1F_1}(i\omega M,1,i\omega \xi)$ can be expanded around $i\omega \xi=\infty$ as
\begin{align}
{_1F_1}(i\omega M,1,i\omega \xi)\approx \xi^{-i\omega M}\left[1+\frac{\omega^2M^2}{\xi} +O\left(\frac{1}{\xi}\right)^2\right]\,.
\end{align}
The above equation can be evaluated efficiently in the numerical processes.


\bsp	
\label{lastpage}
\end{document}